\documentclass[universe,article,accept,pdftex,moreauthors]{mdpi} 

\usepackage{graphicx}
\usepackage{dcolumn}
\usepackage{bm}
\usepackage{amssymb}
\usepackage{latexsym}
\usepackage{booktabs}
\usepackage{amsmath}
\usepackage{multirow}
\usepackage{url}
\usepackage{float}

\newcommand{\ve}{\varepsilon}

\usepackage[normalem]{ulem}
\usepackage{color}
\usepackage{array}
\usepackage{enumerate}

\firstpage{1} 
\pubvolume{1}
\issuenum{1}
\articlenumber{0}
\pubyear{2025}
\copyrightyear{2025}
\externaleditor{Firstname Lastname}
\datereceived{12 December 2024} 
\daterevised{10 February 2025} 
\dateaccepted{4 March 2025} 
\datepublished{ } 
\hreflink{https://\\doi.org/} 

\Title{Revisiting Holographic Dark Energy from the Perspective of Multi-Messenger Gravitational Wave Astronomy: Future Joint Observations with Short Gamma-Ray Bursts}

\TitleCitation{Revisiting Holographic Dark Energy from the Perspective of Multi-Messenger Gravitational Wave Astronomy: Future Joint Observations with Short Gamma-Ray Bursts}

\Author{Tao Han~
 $^{1}$, Ze Li $^{1}$, Jing-Fei Zhang 
 $^{1,}$* and Xin Zhang $^{1,2,3,}$*\orcidA{}}

\AuthorNames{Tao Han, Ze Li, Jing-Fei Zhang and Xin Zhang}

\AuthorCitation{Han, T.; Li, Z.; Zhang, J.-F.; Zhang, X.}

\address{%
$^{1}$ \quad Liaoning Key Laboratory of Cosmology and Astrophysics, College of Sciences, Northeastern University, Shenyang 110819, China; hantao@stumail.neu.edu.cn (T.H.); lizeneu@stumail.neu.edu.cn (Z.L.) 
\\
$^{2}$ \quad MOE Key Laboratory of Data Analytics and Optimization for Smart Industry, Northeastern University, Shenyang 110819, China\\
$^{3}$ \quad National Frontiers Science Center for Industrial Intelligence and Systems Optimization, Northeastern University, Shenyang 110819, China}

\corres{Correspondence: jfzhang@mail.neu.edu.cn (J.-F.Z.); zhangxin@mail.neu.edu.cn (X.Z.)}

\abstract{The advent of third-generation (3G) gravitational-wave (GW) detectors opens new opportunities for multi-messenger observations of binary neutron star merger events, holding significant potential for probing the history of cosmic expansion. In this paper, we investigate the holographic dark energy (HDE) model by using the future GW standard siren data observed from the 3G GW detectors and the short $\gamma$-ray burst THESEUS-like detector joint observations. We find that GW data alone can achieve a relatively precise estimation of the Hubble constant, with precision of $0.2$--$0.6\%$, but its ability to constrain other cosmological parameters remains limited. Nonetheless, since the GW data can break parameter degeneracies generated by the mainstream EM observations, CMB~+~BAO + SN (CBS), GW standard sirens play a crucial role in enhancing the accuracy of parameter estimation. With the addition of GW data to CBS, the constraints on cosmological parameters $H_0$, $c$ and $\Omega_{\rm{m}}$ can be improved by $63$--$88\%$, $27$--$44\%$ and $55$--$70\%$. In summary, observations of GW standard sirens from 3G GW detectors could be pivotal in probing the fundamental nature of dark energy.}


\begin{document}

\section{Introduction}

Observations of type Ia supernovae (SN) revealed that the expansion of the universe is accelerating, which is explained by dark energy with negative pressure in modern cosmology~\cite{Riess:1998cb,Perlmutter:1998np,Sahni:1999gb,Caldwell:1999ew,Padmanabhan:2002ji,Peebles:2002gy,Copeland:2006wr,Li:2011sd,Bamba:2012cp}. To deepen our understanding of the universe, various models have been proposed to investigate the nature of dark energy. Among these models, the $\Lambda$ cold dark matter ($\Lambda$CDM) model is the preferred one, since it fits exceptionally well with the majority of cosmological observations. However, it still suffers from severe theoretical puzzles, namely, the ``fine-tuning'' and ``coincidence'' problems~\cite{Weinberg:1988cp,Carroll:2000fy}. Thus, searching for new physics beyond the $\Lambda$CDM model is an important mission in modern cosmology.

The simplest extension of $\Lambda$CDM cosmology is the model with a dark energy characterized by a constant equation-of-state (EoS) parameter, $w$, commonly referred to as the $w$CDM model. Another interesting attempt is a model of holographic dark energy (HDE)~\cite{Li:2004rb}, which also contains just one extra parameter relative to the $\Lambda$CDM model. Compared to the $w$CDM model, the HDE model is supported by stronger theoretical foundations. It combines the holographic principle of quantum gravity with the effective quantum field theory~\cite{Cohen:1998zx,Li:2004rb} and has been widely investigated in the literature~\cite{Zhang:2005hs,Zhang:2007sh,Huang:2004wt,Wang:2004nqa,Nojiri:2005pu,Chang:2005ph,Zhang:2007an,Li:2012spm,Zhang:2014ija,Landim:2015hqa,Wang:2016och,Li:2010xjz,Xu:2016grp,Li:2017usw,Feng:2018yew,Li:2013dha,Wang:2023gov,Li:2024qus,Nojiri:2017opc,Nojiri:2020wmh,Nojiri:2021iko,Nojiri:2021jxf,Nojiri:2019kkp}. Moreover, it was found that the ``fine-tuning'' and ``coincidence'' problems can be partly resolved in this model~\cite{Li:2004rb}. To date, the HDE model remains a competitive candidate among various dark energy models regarding its ability to fit observational data~\cite{Xu:2016grp}. In addition to the HDE model, another theoretical variant, known as the Ricci dark energy (RDE) model, has been proposed~\cite{Gao:2007ep,Zhang:2009un,Cai:2008nk,Fu:2011ab,Cui:2014sma,Zhang:2014sqa,Yu:2015sla}. This model employs the average radius of the Ricci scalar curvature instead of the universe's future event horizon as the infrared cutoff within the theoretical framework of HDE~\cite{Gao:2007ep,Zhang:2009un}. However, the RDE model is not favored by the current observations~\cite{Xu:2016grp}. 

Recently, Li et al.
~\cite{Li:2024qus} investigated the cosmological implications of the HDE model by utilizing cosmic microwave background (CMB), Dark Energy Spectroscopic Instrument (DESI) 2024 baryon acoustic oscillations (BAO), and SN data. They found that, based on the Bayesian evidence, the HDE model is statistically comparable to the $\Lambda$CDM model when evaluated using DESI BAO data in combination with SN data. However, the inclusion of CMB data makes the HDE model significantly less favored than the $\Lambda$CDM model. Thus, the HDE model warrants additional investigation, especially with the availability of more precise late-universe observations in the future. Furthermore, the Hubble constant inferred from the CMB anisotropies by the Planck 
 mission and the Cepheid-supernova distance ladder measurement are inconsistent, with a more-than-5$\sigma$ discrepancy~\cite{Riess:2021jrx}, which is the so-called ``Hubble tension'' problem (see, e.g., Refs.~\cite{cai:2020,Guo:2019dui,Yang:2018euj,Vagnozzi:2019ezj,DiValentino:2019jae,DiValentino:2019ffd,Liu:2019awo,Zhang:2019cww,Ding:2019mmw,Lin:2020jcb,Hryczuk:2020jhi,Cai:2021wgv,Vagnozzi:2021tjv,Vagnozzi:2021gjh,Riess:2019qba,Verde:2019ivm,Li:2024qso} for related discussions). Therefore, we actually need new cosmological probes to make an arbitration for the Hubble tension. In fact, the gravitational wave (GW) standard siren is one of the most promising cosmological probes.

As proposed by Schutz in 1986~\cite{Schutz:1986gp,Holz:2005df}, the absolute luminosity distance to the source can be
independently obtained by the analysis of the GW waveform. If the redshift of the source can also be obtained by its EM counterpart, then we can establish the true distance-redshift relation to explore the expansion history of the universe and constrain the cosmological parameters such as the Hubble constant~\cite{Zhao:2010sz,Zhang:2018byx,Jin:2020hmc,Li:2019ajo,Cai:2017aea,Cai:2016sby,Zhang:2019ple,Jin:2023sfc,KAGRA:2013rdx,Song:2022siz,Han:2023exn}. Actually, the only GW-EM multi-messenger observation, GW170817, has given the first measurement of the Hubble constant using the standard siren method with a precision of about $14\%$~\cite{LIGOScientific:2017adf}. As a result, the current measurements are far from making an arbitration for the Hubble tension. Therefore, the researchers have to resort to future GW observations.

The third generation (3G) ground-based GW detectors, such as the Einstein Telescope (ET)~\cite{ET-web,Punturo:2010zz} and the Cosmic Explorer (CE)~\cite{CE-web,Evans:2016mbw}, will have more than one order of magnitude improvement over the current detectors~\cite{Evans:2021gyd}. Consequently, in the era of 3G GW detectors, more binary neutron star (BNS) mergers will be detected at much deeper redshifts. Meanwhile, their associated $\gamma$-ray bursts (GRBs) can be accurately localized by GRB detectors. Then, the redshifts can be obtained by optical to Near Infra-Red afterglow spectra (that unambiguously pinpoints the host galaxy) from ground-based follow-up observations~\cite{THESEUS:2021uox}. In this work, we focus on the collaboration between 3G GW detectors and a future GRB detector similar to the proposed Transient High-Energy Sky and Early Universe Surveyor (THESEUS) mission~\cite{THESEUS:2017qvx,THESEUS:2017wvz,Stratta:2018ldl}.

In this paper, we revisit both the HDE and RDE models by considering future GW and GRB joint observations. Compared to previous work~\cite{Zhang:2019ple}, our main highlights in this paper are as follows: (i) We perform a detailed and rigorous analysis of GW-GRB detection, directly calculating the redshift distribution of GW-GRB events rather than assuming 1000~detected standard sirens over a 10-year observation, as adopted in Refs.~\cite{Zhao:2010sz,Zhang:2018byx,Jin:2020hmc,Li:2019ajo,Cai:2017aea,Cai:2016sby,Zhang:2019ple}. (ii)~Since the impact of the Earth's rotation for the 3G GW detectors cannot be ignored~\cite{Han:2023exn}, in our simulation of GW standard sirens, we incorporate Earth's rotational effects to better reflect real observational conditions. (iii) We conduct a cosmological analysis for four different cases of 3G GW observations: single ET, single CE, the CE-CE network, and the ET-CE-CE network, instead of considering only a single ET as in previous work~\cite{Zhang:2019ple}. Additionally, we examine both optimistic and realistic scenarios for the THESEUS field of view (FoV) in GRB detection. Through this comprehensive analysis, we highlight the potential of 3G-era standard sirens in constraining the cosmological parameters of the HDE and RDE models. Note that although the RDE model is disfavored by current observations, we include it as a demonstrative case to illustrate the forecasting potential of GW-GRB joint observations for cosmological parameter estimation.

This paper is organized as follows. In Section~\ref{sec2} we provide a brief description of the HDE and RDE models. In Section~\ref{sec3}, we describe the method to simulate the GW standard siren data. In Section~\ref{sec4}, we report the constraint results of cosmological parameters and make some discussions. In Section~\ref{sec5}, we present a final conclusion.
\section{Cosmological Model}\label{sec2}

In quantum field theory, one of the most enduring mysteries is the stark discrepancy between the theoretical prediction and observational measurement of the vacuum energy density. When the Planck scale is used as an ultraviolet (UV) cutoff, the theoretical estimation of vacuum energy density exceeds the critical density of the universe by an astounding 120 orders of magnitude~\cite{Weinberg:1988cp}. This incongruity highlights the limitations of current physical theories and stems from the absence of a complete theory of quantum gravity.

To address this issue, the HDE model is proposed. This model is grounded in effective quantum field theory and incorporates gravitational effects along with the holographic principle~\cite{tHooft:1993dmi,Susskind:1994vu}. When gravity is taken into account, the number of degrees of freedom in a given spatial region must be constrained, as an excessive number of degrees of freedom could result in the formation of a black hole~\cite{Cohen:1998zx}. As a result, the energy density of the vacuum can be expressed as~\cite{Li:2004rb} 
\begin{equation} 
	\rho_{\rm de}=3c^{2} M_{\rm pl}^{2}L^{-2}, 
\end{equation}
where $L$ is the infrared (IR) cutoff, representing the largest allowable length scale in the effective field theory, $M_{\rm pl}$ denotes the reduced Planck mass, and $c$ is a dimensionless phenomenological parameter that encapsulates uncertainties in theoretical predictions.

This framework effectively shifts the focus from resolving the UV cutoff problem to identifying an appropriate IR cutoff scale. Different choices for the IR cutoff scale give rise to various holographic models of dark energy. In this work, we explore two prominent models: the HDE model and the RDE model.

\subsection{The HDE Model}

The HDE model is defined by choosing the event horizon size of the universe as the IR cutoff~\cite{Li:2004rb}. Thus, the energy density is given by
\begin{equation} 
	\rho_{\rm de}=3c^{2} M_{\rm pl}^{2}R_{\rm eh}^{-2}.
\end{equation}
where  $R_{\rm eh}$ is the future event horizon expressed as
\begin{equation}
	R_{\rm eh}=a\int_{t}^{\infty}\frac{{\rm d}t'}{a}=a\int_{a}^{\infty}\frac{{\rm d}a'}{H(a'){a'}^2}, 
\end{equation}
with $a$ the scale factor of the universe and $H(a)$ the Hubble parameter as a function of $a$.

In the HDE model, the evolution of dark energy density is governed by the following differential equations:
\begin{equation}
	\begin{aligned}
		\frac{1}{E(z)}\frac{{\rm d}E(z)}{{\rm d}z}=-\frac{\Omega_{\rm{de}}(z)}{1+z}\left(\frac{1}{2}+\frac{\sqrt{\Omega_{\rm{de}}(z)}}{c}-\frac{3}{2\Omega_{\rm{de}}(z)}\right),\\
		\frac{{\rm d}\Omega_{\rm{de}}(z)}{{\rm d}z}=-\frac{2\Omega_{\rm{de}}(z)(1-\Omega_{\rm{de}}(z))}{1+z}\left(\frac{1}{2}+\frac{\sqrt{\Omega_{\rm{de}}(z)}}{c}\right),
	\end{aligned}
\end{equation} 
where $E(z)\equiv H(z)/H_0$ is the dimensionless Hubble parameter, and $\Omega_{\rm{de}}(z)$ is the dark energy density fraction. By solving these equations with the initial conditions $E(0)\equiv 1$ and $\Omega_{\rm{de}}(0)\equiv 1-\Omega_{\rm{m}}$, we can determine the evolution of $E(z)$ and $\Omega_{\rm{de}}(z)$. Then from the energy conservation equations,
\begin{equation}\label{eq1}
	\begin{aligned}
		\dot{\rho}_{\rm de} +3H(1+w)\rho_{\rm de}= 0,\\
		\dot{\rho}_{\rm m} +3H\rho_{\rm m}= 0,
	\end{aligned}
\end{equation}
where a dot denotes the derivative with respect to the cosmic time $t$ and $\rho_{\rm m}$ represents the matter density, one can obtain the EoS of dark energy in the HDE model
\begin{equation}\label{eq2}
	w=-\frac{1}{3}-\frac{2\sqrt{\Omega_{\rm{de}}}}{3c}.
\end{equation}
From 
 Equation~(\ref{eq2}), \textls[-15]{we can see that the HDE model cannot reduce to the $\Lambda$CDM model~\cite{Li:2010xjz,Xu:2016grp}.}

\subsection{The RDE Model}

The RDE model adopts the average radius of the Ricci scalar curvature as the IR cutoff~\cite{Gao:2007ep,Zhang:2009un}. In an FRW spatially-flat universe, the Ricci scalar is expressed as
\begin{equation} 
	R=-6(\dot{H}+2H^2).
\end{equation}
Then 
 the dark energy density in the RDE model can be expressed as 
\begin{equation} 
	\rho_{\rm de}=3\gamma M_{\rm pl}^{2}(\dot{H}+2H^2), 
\end{equation}
where $\gamma$ is a positive constant that can be redefined in terms of the phenomenological parameter $c$.

The evolution equation for the RDE model is
\begin{equation} 
	E^2=\Omega_{\rm m}e^{-3x}+\gamma\left(\frac{1}{2}\frac{{\rm d}E^2}{{\rm d}x}+2E^2\right), 
\end{equation} 
where $x=\ln a$. Solving this differential equation, we obtain
\begin{equation}
	E(z)=\bigg(\frac{2\Omega_{\rm m}}{2-\gamma}(1+z)^{3}+\bigg(1-\frac{2\Omega_{\rm m}}{2-\gamma}\bigg)(1+z)^{(4-\frac{2}{\gamma})}\bigg)^{1/2}.
\end{equation}
Furthermore, 
 from the energy conservation Equation (\ref{eq1}), we can obtain the EoS of dark energy in the RDE model
\begin{equation}
	w=\frac{\frac{\gamma-2}{3\gamma}f_0e^{-(4-\frac{2}{\gamma})x}}{\frac{\gamma}{2-\gamma}\Omega_{\rm m}e^{-3x}+f_0e^{-(4-\frac{2}{\gamma})x}},
\end{equation}
 where $f_0$ is determined as
\begin{equation}
	f_0=1-\frac{2}{2-\gamma}\Omega_{\rm m}.
\end{equation}
Same 
 as the HDE model, the RDE model cannot reduce to the $\Lambda$CDM model~\cite{Li:2010xjz,Xu:2016grp}.

\section{Methodology}\label{sec3}

\subsection{Simulations of GW Events}

In this subsection, we briefly review the simulation of BNS mergers for the following analysis. To create a catalog of these events, we need to obtain the redshift distribution of their mergers. Using the star formation rate~\cite{Vitale:2018yhm,Belgacem:2019tbw,Yang:2021qge}, we define the merger rate density in the observer frame as
\begin{equation}
R_{\rm m}(z)=\frac{\mathcal{R}_{\rm m}(z)}{1+z} \frac{{\rm d}V(z)}{{\rm d}z},
\end{equation}
where ${\rm d}V/{\rm d}z$ is the comoving volume element, and $\mathcal{R}_{\rm m}$ represents the merger rate in the source frame, which is given by
\begin{equation}
\mathcal{R}_{\rm m}(z)=\int_{t_{\rm min}}^{t_{\rm max}} \mathcal{R}_{\rm f}[t(z)-t_{\rm d}] P(t_{\rm d}){\rm d}t_{\rm d}.
\end{equation}
Here 
 $\mathcal{R}_{\rm f}$ is simply proportional to the Madau-Dickinson star formation rate~\cite{Madau:2014bja}, $t(z)$ is the age of the universe at the time of merger, $t_{\rm d}$ represents the delay time between BNS system formation and merger, $t_{\rm min}= 20$~Myr is the minimum delay time, $t_{\rm max}$ is maximum delay time taken as the Hubble time~\cite{Vitale:2018yhm}, $P(t_{\rm d})$ is the time delay distribution and we assume the power-law delay model~\cite{Virgili:2009ca,DAvanzo:2014urr}, which is given by $P(t_{\rm d})=1/t_{\rm d}$, with $t_{\rm d}>t_{\rm min}$.

In this paper, we consider the local comoving merger rate to be $\mathcal{R}_{\rm m}(z=0) = \rm 920~Gpc^{-3}~yr^{-1}$, as estimated from the O1 LIGO and
the O2 LIGO/Virgo observation run~\cite{LIGOScientific:2018mvr}, and it is also consistent with the O3 observation run~\cite{KAGRA:2021duu}. We generate a 10-year catalog of BNS mergers. For each merger event, the location $(\theta,\phi)$, orientation angle $\iota$, polarization angle $\psi$, and coalescence phase $\psi_{\rm c}$ are all drawn from uniform distributions. For the masses of neutron stars, we assume that they follow a Gaussian distribution, centered at $1.33~M_{\odot}$ with a standard deviation of $0.09~M_{\odot}$~\cite{LIGOScientific:2018mvr}.

\subsection{Detection of GW Events}

Using the stationary phase approximation (SPA), the frequency-domain GW waveform for a network of $N$ independent detectors can be expressed as 
~\cite{Wen:2010cr}
\begin{equation}
\tilde{\boldsymbol{h}}(f)=e^{-i\boldsymbol\Phi}\boldsymbol h(f),
\end{equation}
where $\boldsymbol\Phi$ is an $N\times N$ diagonal matrix defined as $\Phi_{ij}=2\pi f\delta_{ij}(\boldsymbol{n\cdot r}_k)$, $\boldsymbol{n}$ represents the GW signal's propagation direction and $\boldsymbol{r}_k$ is the spatial location of the $k$-th detector. The $\boldsymbol h(f)$ is defined as
\begin{equation}
\begin{aligned}
\boldsymbol h(f)=&\left [\frac{h_1(f)}{\sqrt{S_{\rm {n},1}(f)}}, \frac{h_2(f)}{\sqrt{S_{\rm {n},2}(f)}},\ldots,\frac{h_N(f)}{\sqrt{S_{{\rm n},N}(f)}}\right ]^{\rm T}.
\end{aligned}
\end{equation}
where $S_{{\rm n},k}(f)$ is the one-side noise power spectral density of the $k$-th detector.

In this paper, we focus on the inspiral phase waveform of a non-spinning BNS system. We employ the restricted Post-Newtonian approximation up to 3.5 PN order~\cite{Cutler:1992tc,Sathyaprakash:2009xs}. The frequency-domain GW waveform for the $k$-th detector is given by~\cite{Sathyaprakash:2009xs}
\begin{align}
	h_k(f)=\mathcal A_k f^{-7/6}{\rm exp}
	\{i[2\pi f t_{\rm c}-\pi/4-2\psi_c+2\Psi(f/2)]-\varphi_{k,(2,0)}\},
\end{align}
where $\Psi(f/2)$ and $\varphi_{k,(2,0)}$ are described in Refs.~\cite{Zhao:2017cbb,Blanchet:2004bb}. The Fourier amplitude $\mathcal A_k$ is given~by
\begin{align}
	\mathcal A_k=\frac{1}{d_{\rm L}}\sqrt{(F_{+,k}(1+\cos^{2}\iota))^{2}+(2F_{\times,k}\cos\iota)^{2}}\sqrt{5\pi/96}\pi^{-7/6}\mathcal M^{5/6}_{\rm chirp},
\end{align}
where $d_L$ is the luminosity distance, $F_{+,k}$, $F_{\times,k}$ are the antenna response functions, $\mathcal M_{\rm chirp}$ is the chirp mass, $M=m_1+m_2$ is the total mass of the binary, and $\eta=m_1 m_2/M^2$ is the symmetric mass ratio. Under the SPA, $F_{+,k}$, $F_{\times,k}$ and $\Phi_{ij}$ are functions of frequency, with time t replaced by $t_{\rm f}=t_{\rm c}-(5/256)\mathcal M^{-5/3}_{\rm chirp}(\pi f)^{-8/3}$, where $t_{\rm c}$ is the coalescence time~\cite{Maggiore:2007ulw}.

In this work, we determine that a GW event is detectable only if its signal-to-noise ratio (SNR) exceeds a threshold of 12. For low-mass systems, the combined SNR for the detection network of $N$ independent detectors is
\begin{equation}
\rho=(\tilde{\boldsymbol h}|\tilde{\boldsymbol h})^{1/2},     
\end{equation}
where the inner product is defined as
\begin{equation}
(\boldsymbol a|\boldsymbol b)=2\int_{f_{\rm lower}}^{f_{\rm upper}}\{\boldsymbol a(f)\boldsymbol b^*(f)+\boldsymbol b(f)\boldsymbol a^*(f)\}{\rm d}f,
\end{equation}
with $\boldsymbol a$ and $\boldsymbol b$ being the column matrices of the same dimension and * denoting the conjugate transpose. The lower cutoff frequency is $f_{\rm lower}=1$ Hz for ET and $f_{\rm lower}=5$ Hz for CE. $f_{\rm upper}=2/(6^{3/2}2\pi M_{\rm obs})$ is the frequency at the last stable orbit, with $M_{\rm obs}=(m_1+m_2)(1+z)$. 

\subsection{Detection of Short GRBs}

A GRB is detectable only if its peak flux exceeds the detector's sensitivity threshold. Observations such as GW170817/GRB170817A suggest that short GRBs are consistent with a Gaussian-shaped jet profile model~\cite{Howell:2018nhu}
\begin{equation}
	L_{\rm iso}(\theta_{\rm v})=L_{\rm on}{\rm exp}\left(-\frac{\theta^2_{\rm v}}{2\theta^2_{\rm c}} \right),
	\label{eq:jet}
\end{equation}
where $L_{\rm iso}(\theta_{\rm v})$ is the isotropically equivalent luminosity per unit solid angle, $\theta_{\rm v}$ is the viewing angle, $L_{\rm on}$ represents the on-axis isotropic luminosity defined as $L_{\rm on} \equiv L_{\rm iso}(0)$, and $\theta_{\rm c}=4.7^{\circ}$ is the characteristic core angle. We assume the jets align with the binary's orbital angular momentum, meaning $\iota=\theta_{\rm v}$.

The probability of detecting a short GRB depends on distribution $\Phi(L){\rm d}L$, where $\Phi(L)$ is the intrinsic luminosity function and $L$ is the peak luminosity in the 1--10,000 keV energy range in the rest frame assuming isotropic emission. We adopt a broken-power-law luminosity model for $\Phi(L)$
\begin{equation}
	\Phi(L)\propto
	\begin{cases}
		(L/L_*)^{\alpha_{\rm L}}, & L<L_*, \\
		(L/L_*)^{\beta_{\rm L}}, & L\ge L_*,
	\end{cases}
	\label{eq:distribution}
\end{equation}
where $L_{*}$ is the characteristic luminosity separating the two regimes, with power-law slopes $\alpha_{\rm L}$ and $\beta_{\rm L}$ for each region. Based on Ref.~\cite{Wanderman:2014eza}, we use $\alpha_{\rm L}=-1.95$, $\beta_{\rm L}=-3$, and $L_{*}=2\times10^{52}$ erg s$^{-1}$. We treat $L_{\rm on}$ as the peak luminosity $L$ and apply a standard low end cutoff of $L_{\rm min} = 10^{49}$ erg s$^{-1}$.

To assess the detectability of a GRB, we need to convert the GRB satellite's flux limit $P_{\rm T}$ to isotropic-equivalent luminosity $L_{\rm iso}$. According to the flux-luminosity relationship for GRBs~\cite{Meszaros:1995dj,Meszaros:2011zr}, the conversion is given by 
\begin{equation}
	L_{\rm iso}=4\pi d_{\rm L}^2(z)k(z)C_{\rm det}/(1+z)P_{\rm T},
\end{equation}
where $C_{\rm det}$ and $k(z)$ are detailed in Refs.~\cite{Howell:2018nhu,Wanderman:2014eza,Band:2002te}. Finally, we use Equation~(\ref{eq:distribution}) to select detectable GRBs from the BNS sample by sampling $\Phi(L){\rm d}L$.

For the THESEUS mission~\cite{Stratta:2018ldl}, a GRB detection is recorded if the value of observed flux is greater than the flux threshold $P_{\rm T}=0.2~\rm ph~s^{-1}~cm^{-2}$ in the 50--300 keV band. We also assume a duty cycle of 80\% and a sky coverage fraction of 0.5. Note that the X-$\gamma$ ray Imaging Spectrometer (XGIS) instrument on THESEUS can localize the source to approximately 5 arcminutes, but this level of precision is achievable only within the central 2 steradians of its FoV. Outside this central area, the localization capability becomes significantly less accurate~\cite{THESEUS:2017qvx,THESEUS:2017wvz,Stratta:2018ldl}. Therefore, in this paper, we consider two scenarios. In the first scenario, termed ``optimistic'', it is assumed that all short GRBs detected by XGIS provide precise redshift estimates through follow-up observations. In contrast, the second scenario, termed ``realistic'', assumes that only approximately one-third of the short GRBs will provide accurate redshift estimates through follow-up observations.

\subsection{Fisher Information Matrix and Error Analysis}

We estimate the instrumental error $\sigma_{d_{\mathrm{L}}}^{\mathrm{inst}}$ in the luminosity distance $d_{\mathrm{L}}$ using the Fisher information matrix. For a network of GW detectors, the Fisher matrix is defined as
\begin{equation}
	F_{ij}=\left(\frac{\partial \tilde{\boldsymbol{h}}}{\partial \theta_i}\Bigg |\frac{\partial \tilde{\boldsymbol{h}}}{\partial \theta_j}\right),
\end{equation}
where $\boldsymbol\theta$ represents nine GW source parameters ($d_{\rm L}$, $\mathcal{M}_{\rm chirp}$, $\eta$, $\theta$, $\phi$, $\iota$, $t_{\rm c}$, $\psi_{\rm c}$, $\psi$) for each event. The instrumental error of GW parameter $\theta _i$ is $\Delta\theta_i=\sqrt{(F^{-1})_{ij}}$.

Additionally, we account for errors due to weak lensing $\sigma_{d_{\mathrm{L}}}^{\mathrm{lens}}$ and peculiar velocity $\sigma_{d_{\mathrm{L}}}^{\mathrm{pv}}$. The weak lensing error is given in Refs.~\cite{Hirata:2010ba,Tamanini:2016zlh,Speri:2020hwc},
\begin{align}
	\sigma_{d_{\rm L}}^{\rm lens}(z)=&\left[1-\frac{0.3}{\pi/2} \arctan(z/0.073)\right]\times d_{\rm L}(z)\nonumber\\ &\times 0.066\left [\frac{1-(1+z)^{-0.25}}{0.25}\right ]^{1.8}.\label{lens}
\end{align}
The 
 error from the peculiar velocity of the GW source follows Ref.~\cite{Kocsis:2005vv}, 
\begin{equation}
	\sigma_{d_{\rm L}}^{\rm pv}(z)=d_{\rm L}(z)\times \left [ 1+ \frac{c(1+z)^2}{H(z)d_{\rm L}(z)}\right ]\frac{\sqrt{\langle v^2\rangle}}{c},\label{pv}
\end{equation}
where $H(z)$ is the Hubble parameter, $c$ is the speed of light, and $\sqrt{\langle v^2\rangle}$ represents the source's peculiar velocity, set to $\sqrt{\langle v^2\rangle}=500\ {\rm km\ s^{-1}}$~\cite{He:2019dhl}.

The total error in $d_{\rm L}$ combines these contributions
\begin{equation}
	\left(\sigma_{d_{\mathrm{L}}}\right)^{2}=\left(\sigma_{d_{\mathrm{L}}}^{\mathrm{inst}}\right)^{2}+\left(\sigma_{d_{\mathrm{L}}}^{\text {lens }}\right)^{2}+\left(\sigma_{d_{\mathrm{L}}}^{\text {pv }}\right)^{2}.\label{eq:total}
\end{equation}

For the GW-GRB observations with $N$ data point, we maximize the likelihood $\mathcal{L}\propto(-\chi^2/2)$ to infer the posterior probability distributions of cosmological parameters $\vec{\Omega}$. The $\chi^2$ function is given by
\begin{align}
	\chi^2=\sum\limits_{i=1}^{N}\left[\frac{{d}_{\rm L}^i-d_{\rm L}({z}_i;\vec{\Omega})}{{\sigma}_{d_{\rm L}}^i}\right]^2,
\end{align}
where ${z}_i$, ${d}_{\rm L}^i$, and ${\sigma}_{d_{\rm L}}^i$ are the $i$-th GW event's redshift, luminosity distance, and the total error of the luminosity distance, respectively.

\section{Results and Discussion}\label{sec4}

In this section, we present the results of the cosmological parameter constraints and provide related discussions. We consider the HDE and RDE models. Firstly, we constrain these cosmological models with GW standard siren data alone. Then, we give the constraint results of CMB+BAO+SN dataset (CBS) and CBS + GW for these cosmological models to show the capability of GW standard sirens for breaking the cosmological parameter degeneracies. For the CMB data, we adopt the ``Planck distance priors'' from the Planck 2018 observation~\cite{Chen:2018dbv}. For the BAO data, we employ the measurements from 6dFGS ($z_{\rm eff}=0.106$)~\cite{Beutler:2011hx}, SDSS-MGS ($z_{\rm eff}=0.15$)~\cite{Ross:2014qpa}, and BOSS DR12 ($z_{\rm eff}=0.38$, 0.51, and 0.61)~\cite{BOSS:2016wmc}. For the SN data, we utilize the latest Pantheon+ compilation without SH0ES subset~\cite{Brout:2022vxf}. For GW data, to ensure a consistent comparison of constraints within the same parameter space, we adopt the best-fit cosmological parameters derived from the CBS data combination as fiducial values for simulating GW data in each cosmological model.

To illustrate the impact of simulated GW data on cosmological parameters, we analyze various configurations of 3G GW detectors: a single ET detector, a single CE detector, a CE-CE network (consisting of a 40-km-arm CE in the United States and a 20-km-arm CE in Australia, referred to as 2CE), and an ET-CE-CE network (comprising one ET detector and two CE-like detectors, referred to as ET2CE). The sensitivity curves for the ET and CE detectors are adopted from Refs.~\cite{ETcurve-web,CEcurve-web}, respectively, as illustrated in Figure~1 
 of Ref.~\cite{Han:2023exn}. Considering the significant uncertainty in the duty cycle of GW detectors, we assume an idealized scenario where all detectors operate with a 100\% duty cycle, following the discussion in Ref.~\cite{Zhu:2021ram}. The geometric parameters characterizing the GW detectors, including latitude $\varphi$, longitude $\lambda$, opening angle $\zeta$, and arm bisector angle $\gamma$ are listed in Table~{\uppercase\expandafter{\romannumeral 1}} of Ref.~\cite{Han:2023exn}. For GRB detection, we consider optimistic and realistic scenarios for the FoV of THESEUS to make our cosmological analysis.

\begin{figure}[H]
	\includegraphics[width=0.9\linewidth,angle=0]{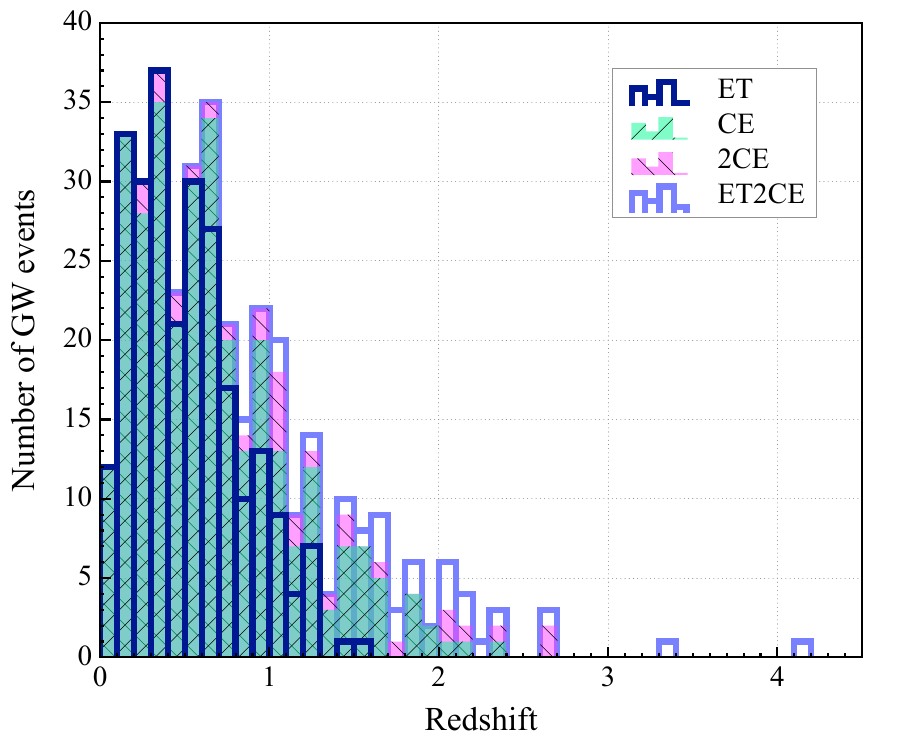}
	\caption{\label{fig1} Redshift 
 distributions of BNS detected by THESEUS in synergy with ET, CE, 2CE, and ET2CE for a 10-year observation in the optimistic scenario.}
\end{figure}

\begin{figure}[H]
	\includegraphics[width=0.9\linewidth,angle=0]{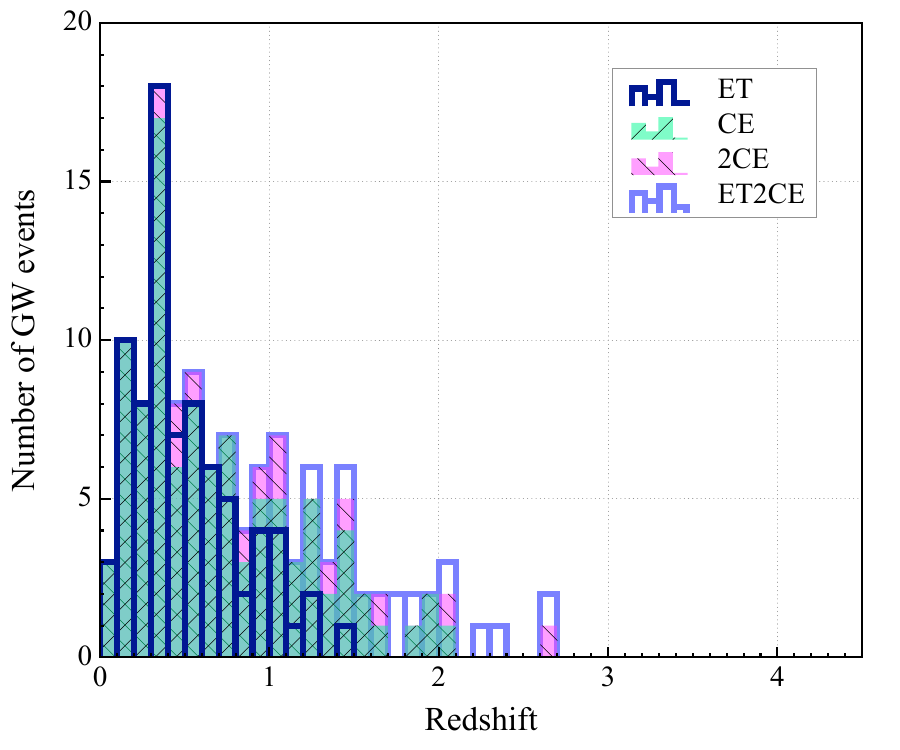}
	\caption{\label{fig2} Same 
 as Figure~\ref{fig1}, but assuming the realistic scenario.}
\end{figure}

\begin{figure}[H]
	\includegraphics[width=0.9\linewidth,angle=0]{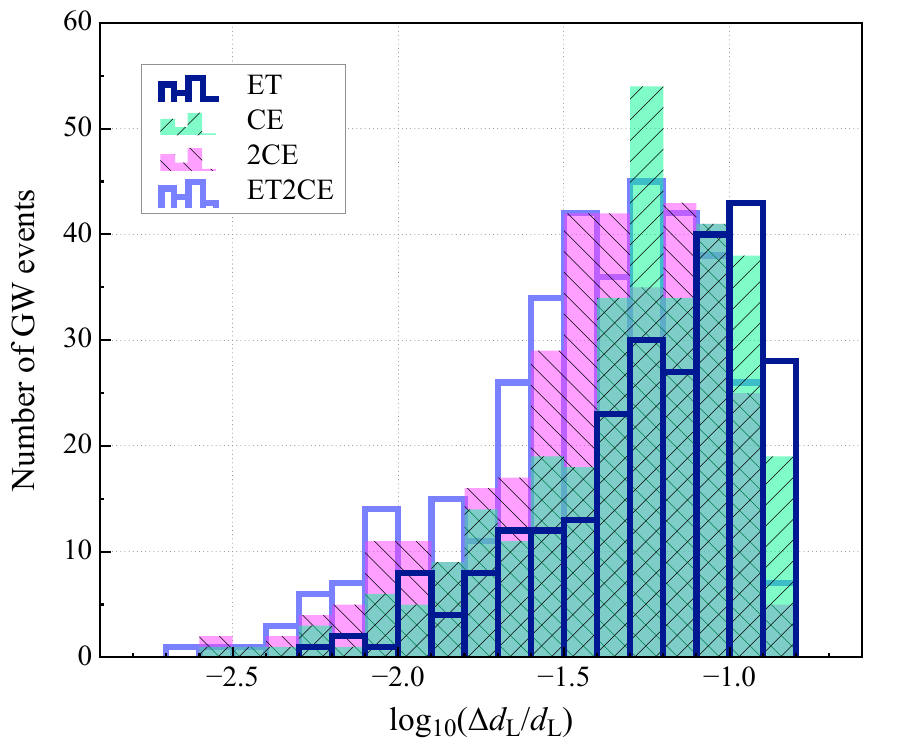}
	\caption{\label{fig3} Distributions 
 of luminosity distance uncertainty $\Delta d_{\rm L}/d_{\rm L}$ of GW standard sirens for ET, CE, 2CE, and ET2CE in the optimistic scenario under the HDE model.}
\end{figure}

\begin{figure}[H]
	\includegraphics[width=0.9\linewidth,angle=0]{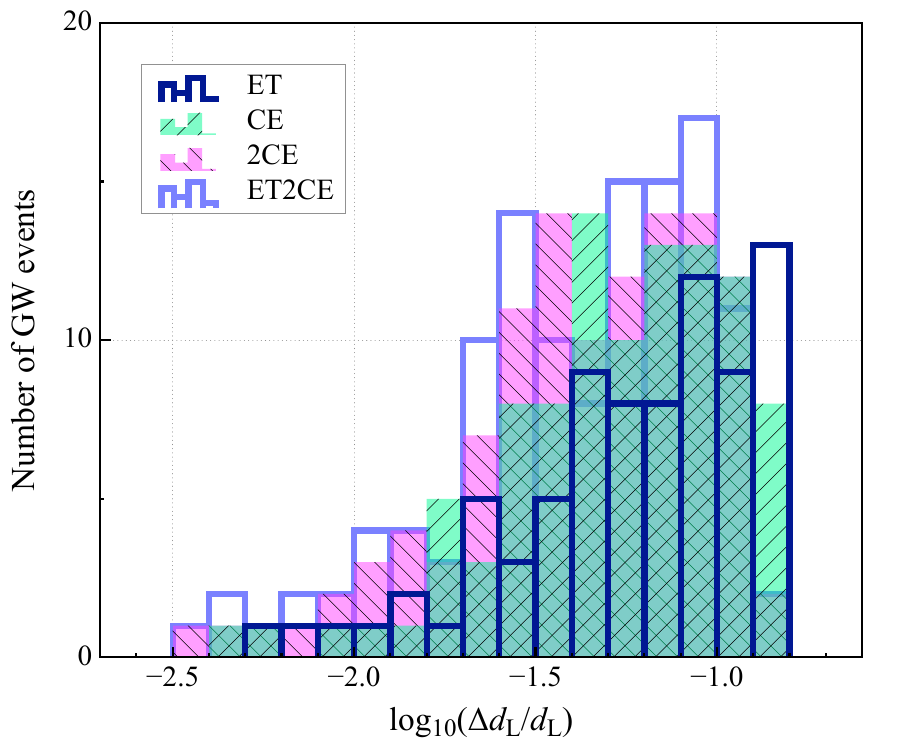}
	\caption{\label{fig4} Same 
 as Fig~\ref{fig3}, but assuming the realistic scenario.}
\end{figure}

{It is worth noting that, in this paper, the parameter settings and computational methods remain consistent with those in our previous work~\cite{Han:2023exn}, except for the choice of the time delay distribution $P(t_{\rm d})$. We adopt the more commonly used power-law delay model for the time delay distribution~\cite{Virgili:2009ca,DAvanzo:2014urr}, which differs from the exponential time delay model employed in our previous work~\cite{Han:2023exn}. This power-law model leads to a relatively conservative estimate of the number of standard sirens (see Ref.~\cite{Yang:2021qge} for a detailed discussion). For completeness, we present the number of standard sirens in Table~\ref{tab1}, their redshift distributions in Figures~\ref{fig1} and~\ref{fig2}, and their luminosity distance uncertainty distributions in Figures~\ref{fig3}~and~\ref{fig4}  for ET, CE, 2CE, and ET2CE under the optimistic and realistic scenarios, as these are key factors influencing the parameter constraints. For more detailed simulation results and related discussions, see Ref.~\cite{Han:2023exn}. From Figures~\ref{fig3} and~\ref{fig4}, we can see that the measurement precision of luminosity distance is about $4$--$13\%$. ET2CE achieves the best measurement precision of $d_{\rm L}$, followed by 2CE, CE, and ET.}

\begin{table}[H]
	\caption{Numbers 
 of GW standard sirens in cosmological analysis, triggered by THESEUS assuming the optimistic and realistic scenarios in synergy with ET, CE, 2CE, and ET2CE, respectively.}
	\label{tab1} 
	\centering
\setlength{\tabcolsep}{11pt} 
\begin{tabular}{c*{4}{>{\centering\arraybackslash}m{1.75cm}}}\toprule
%
		\textbf{Detection Strategy} &\textbf{ET}&\textbf{CE}&\textbf{2CE}&\textbf{ET2CE}\\ \midrule
		Optimistic scenario&      252      &   309   &    340  &  363  \\ 
		Realistic scenario   &      79      &    99  &    112   &  121 \\   \bottomrule
	\end{tabular}
\end{table}

\begin{figure}[H]
	\includegraphics[width=0.45\linewidth,angle=0]{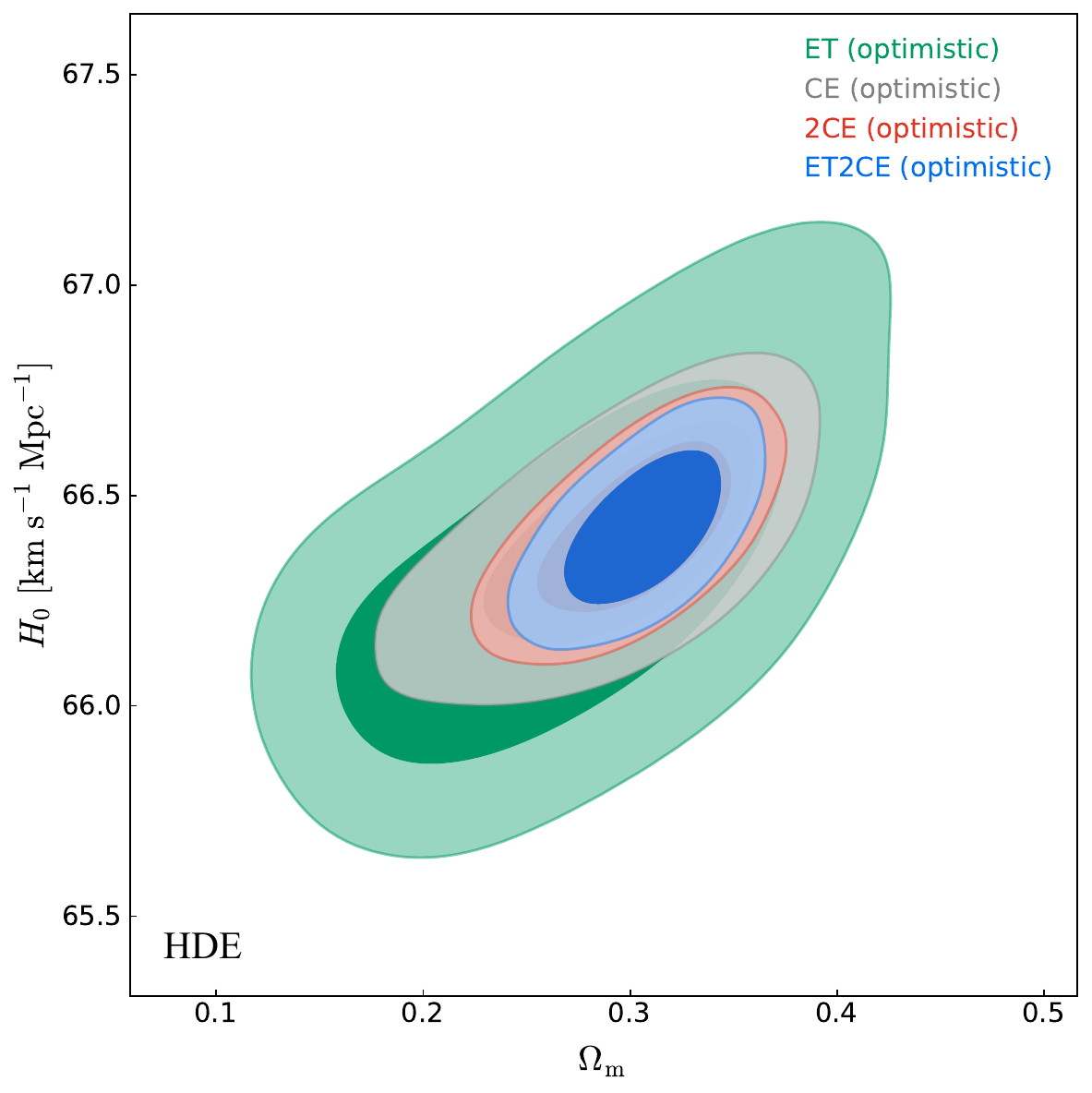}
	\includegraphics[width=0.45\linewidth,angle=0]{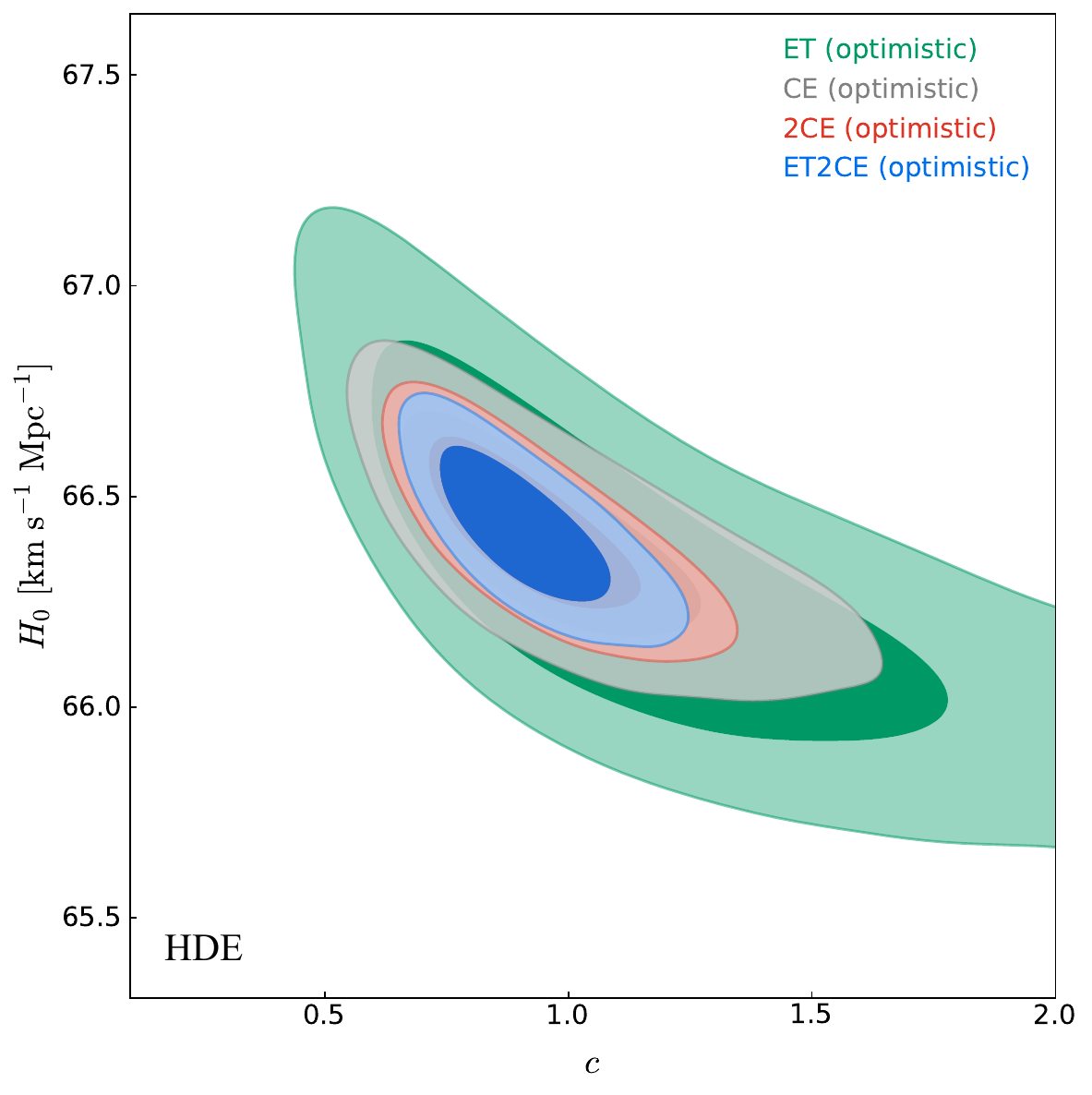}
	\caption{\label{fig5} Two-dimensional 
 marginalized contours ($68.3\%$ and $95.4\%$ confidence level) in the $\Omega_{\rm m}$--$H_0$ and $c$--$H_0$ planes for the HDE model in the optimistic scenario using ET, CE, 2CE and ET2CE, respectively.}
\end{figure}

\begin{figure}[H]
	\includegraphics[width=0.45\linewidth,angle=0]{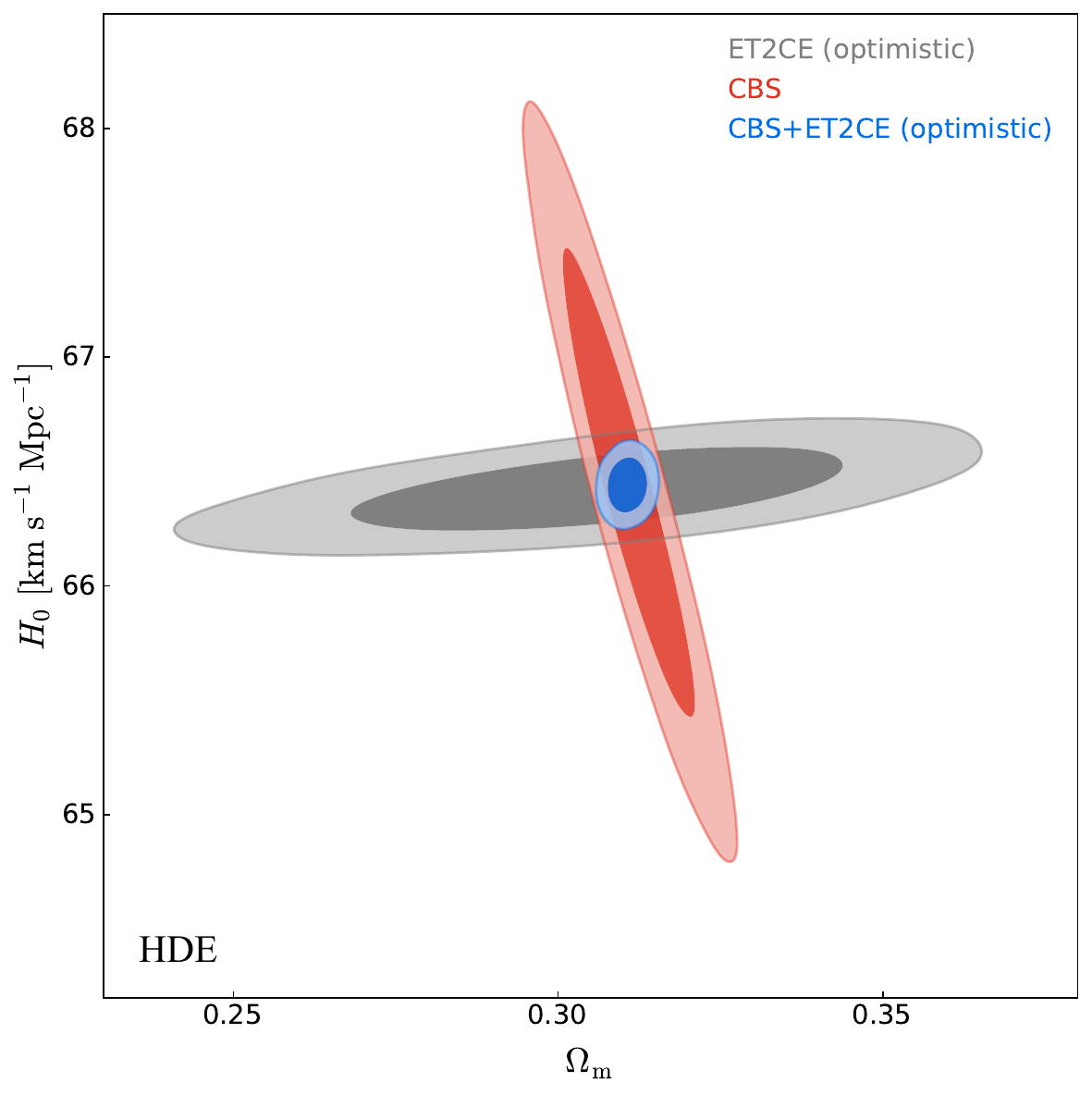}
	\includegraphics[width=0.45\linewidth,angle=0]{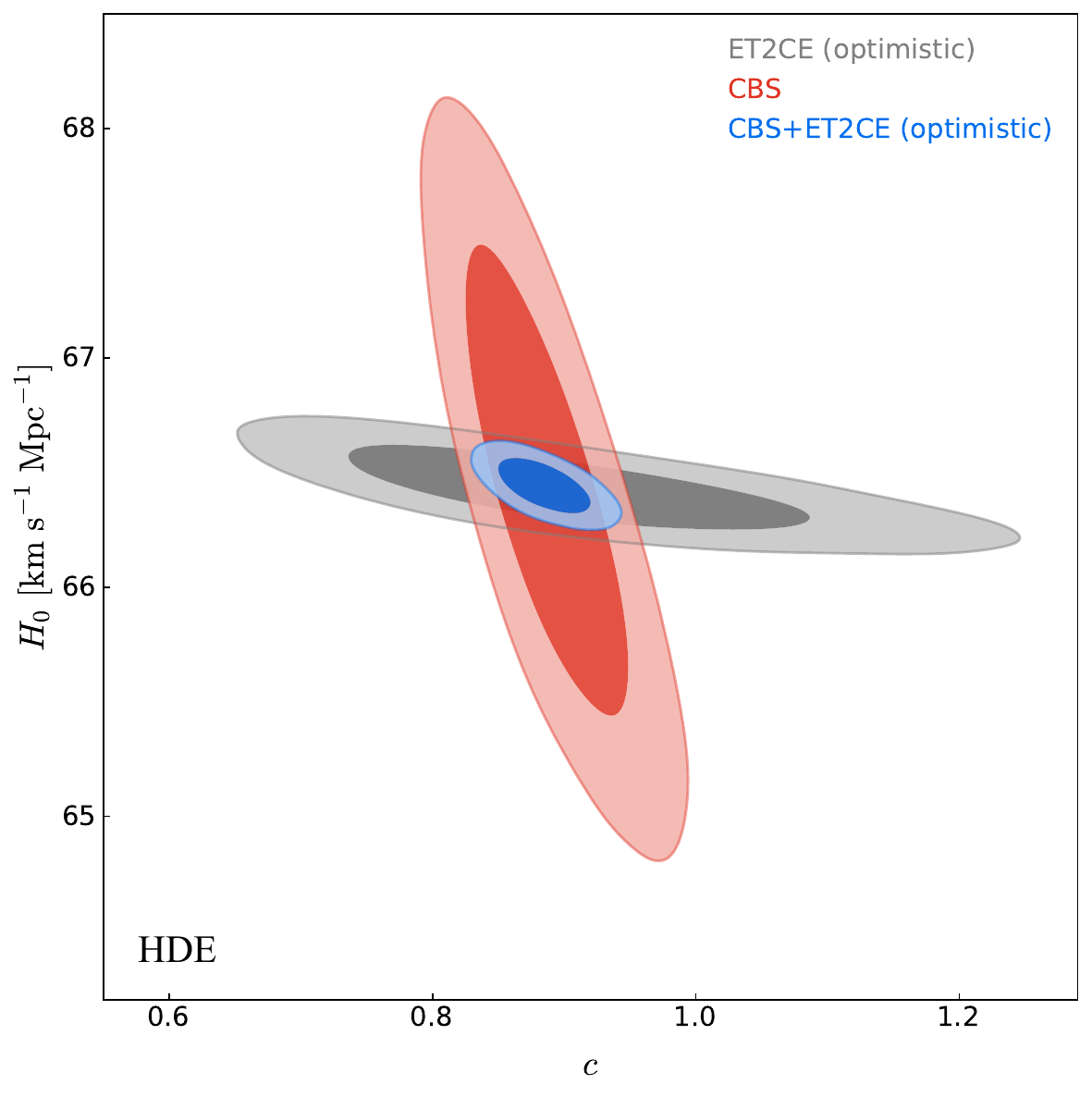}
	\caption{\label{fig6} Two-dimensional 
 marginalized contours ($68.3\%$ and $95.4\%$ confidence level) in the $\Omega_{\rm m}$--$H_0$ and $c$--$H_0$ planes for the HDE model in the optimistic scenario using ET2CE, CBS and CBS + ET2CE data, respectively.}
\end{figure}

\begin{figure}[H]
	\includegraphics[width=0.45\linewidth,angle=0]{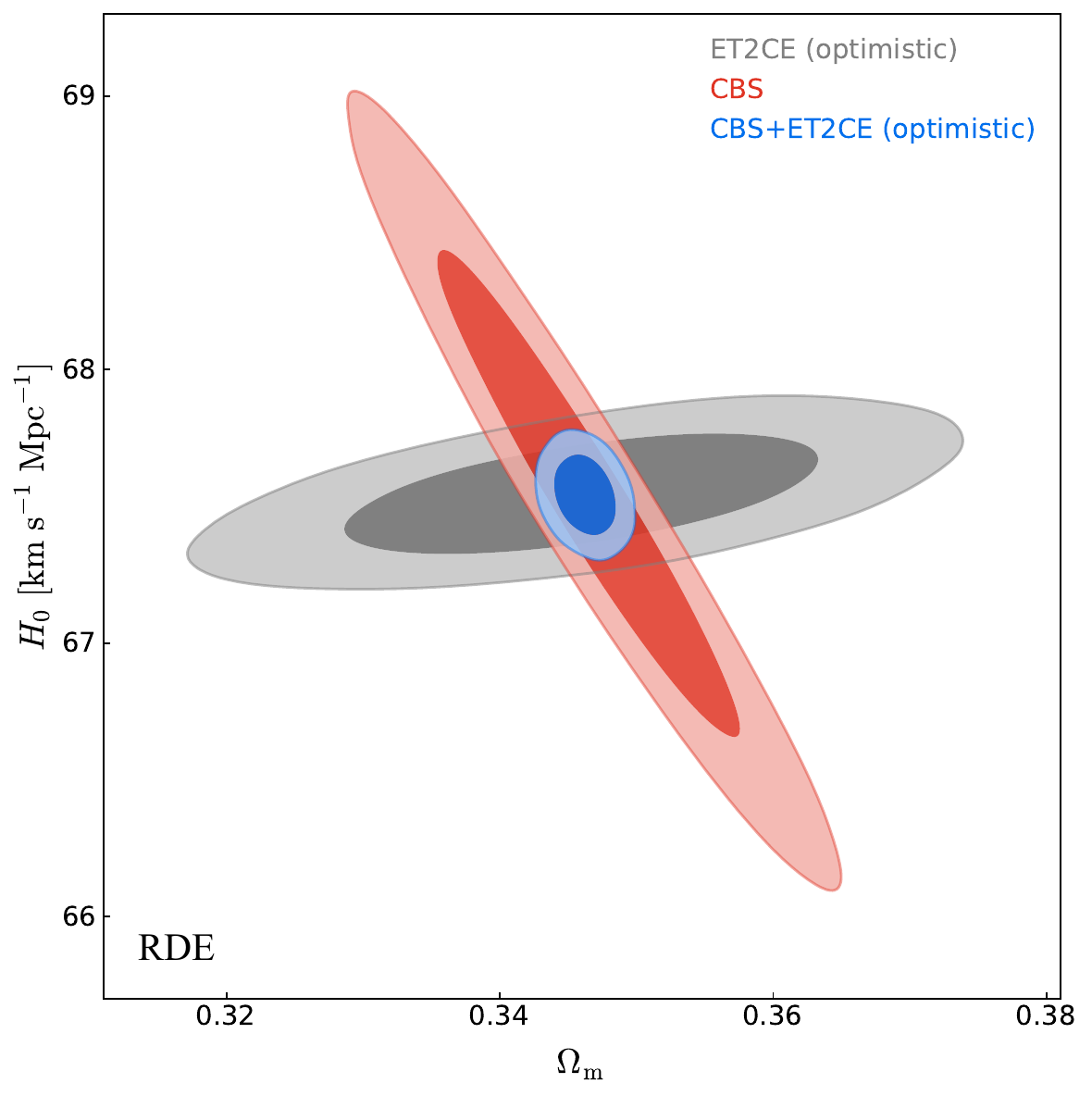}
	\includegraphics[width=0.45\linewidth,angle=0]{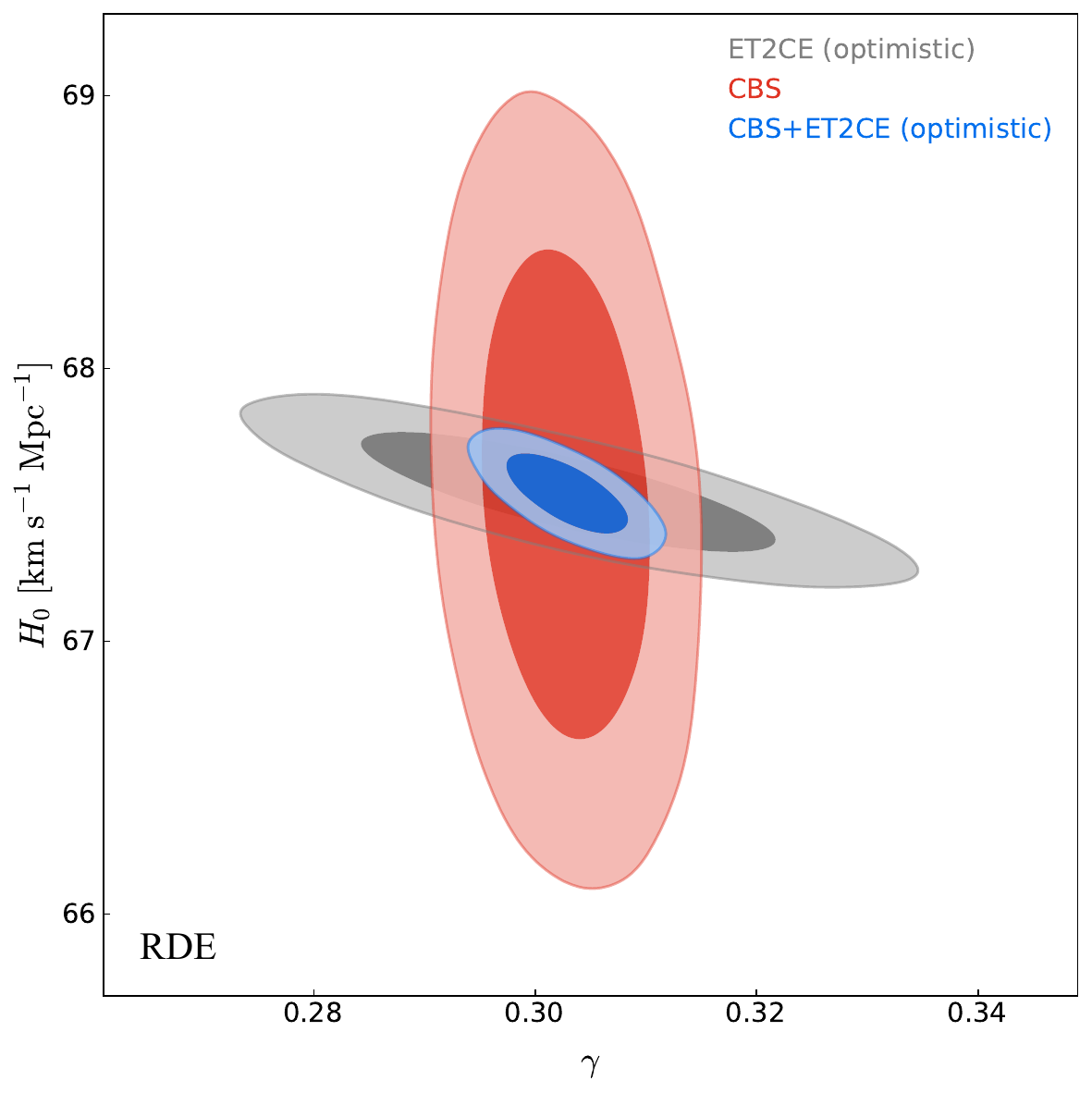}
	\caption{\label{fig7} Two-dimensional 
 marginalized contours ($68.3\%$ and $95.4\%$ confidence level) in the $\Omega_{\rm m}$--$H_0$ and $\gamma$--$H_0$ planes for the RDE model in the optimistic scenario using ET2CE, CBS and CBS + ET2CE data, respectively.}
\end{figure}

\begin{figure}[H]
	\includegraphics[width=0.45\linewidth,angle=0]{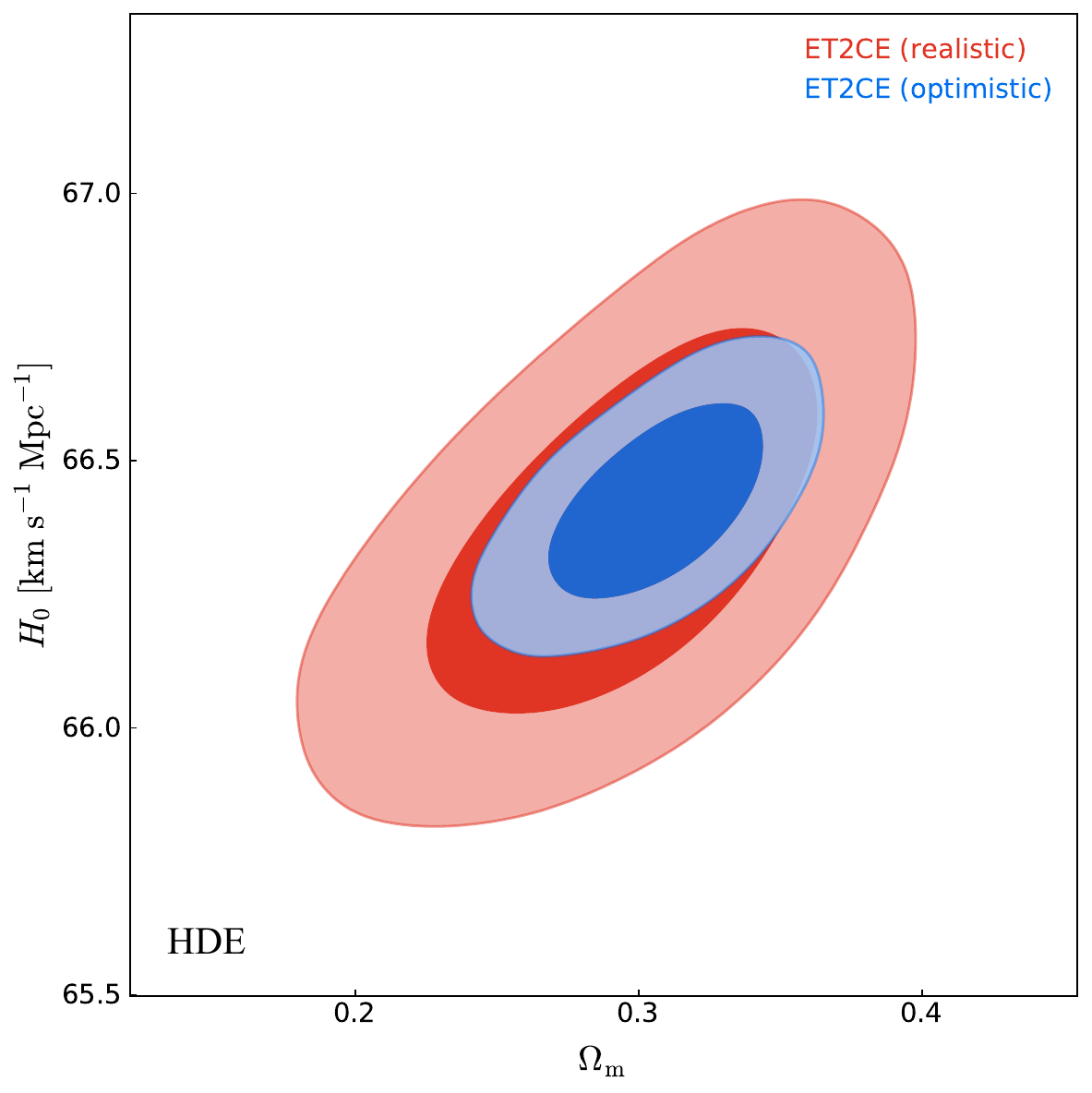}
	\includegraphics[width=0.45\linewidth,angle=0]{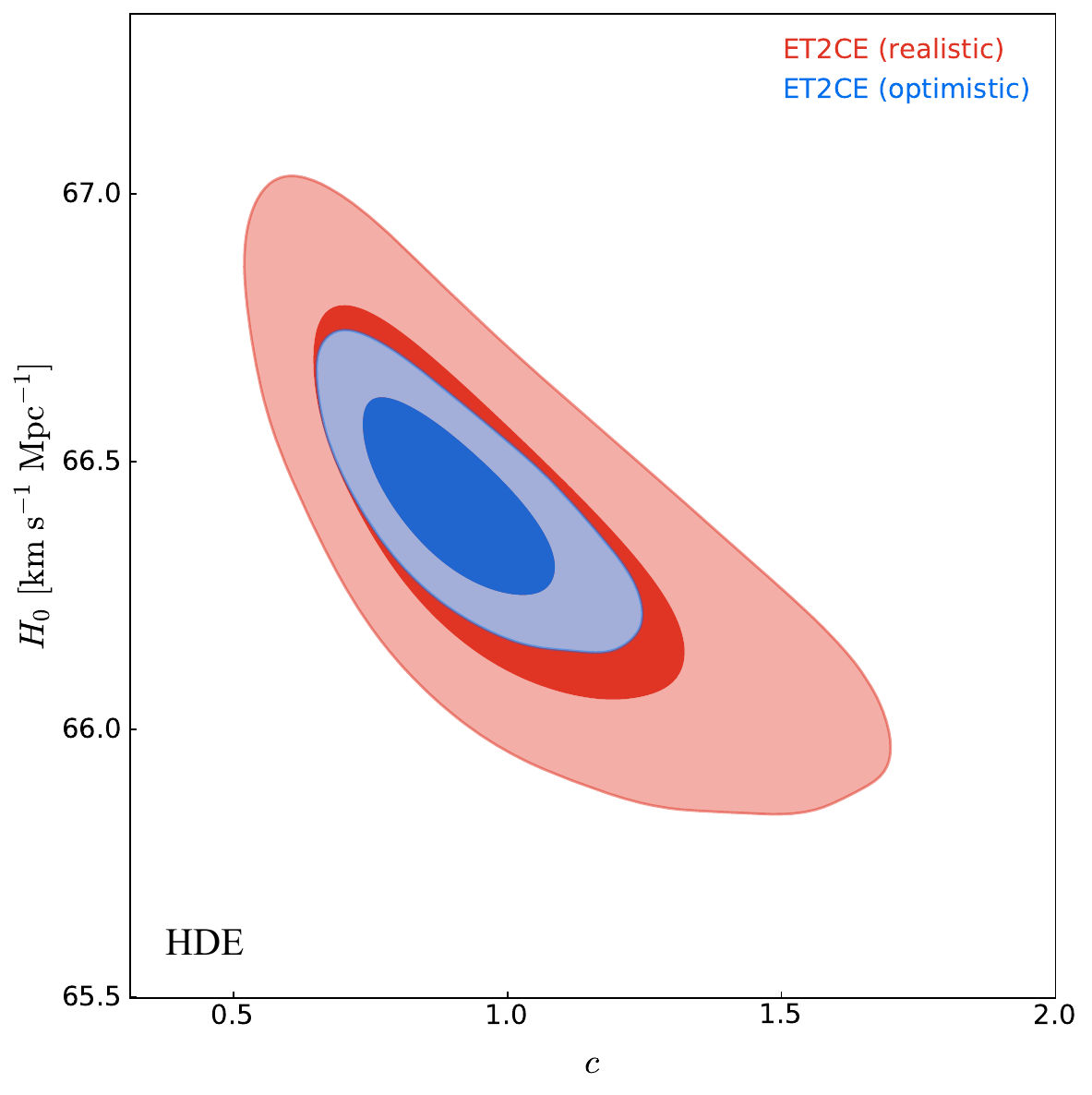}
	\caption{\label{fig8} Two-dimensional 
 marginalized contours ($68.3\%$ and $95.4\%$ confidence level) in the $\Omega_{\rm m}$--$H_0$ and $c$--$H_0$ planes for the HDE model in the realistic and optimistic scenarios of the ET2CE data.}
\end{figure}

\begin{figure}[H]
	\includegraphics[width=0.45\linewidth,angle=0]{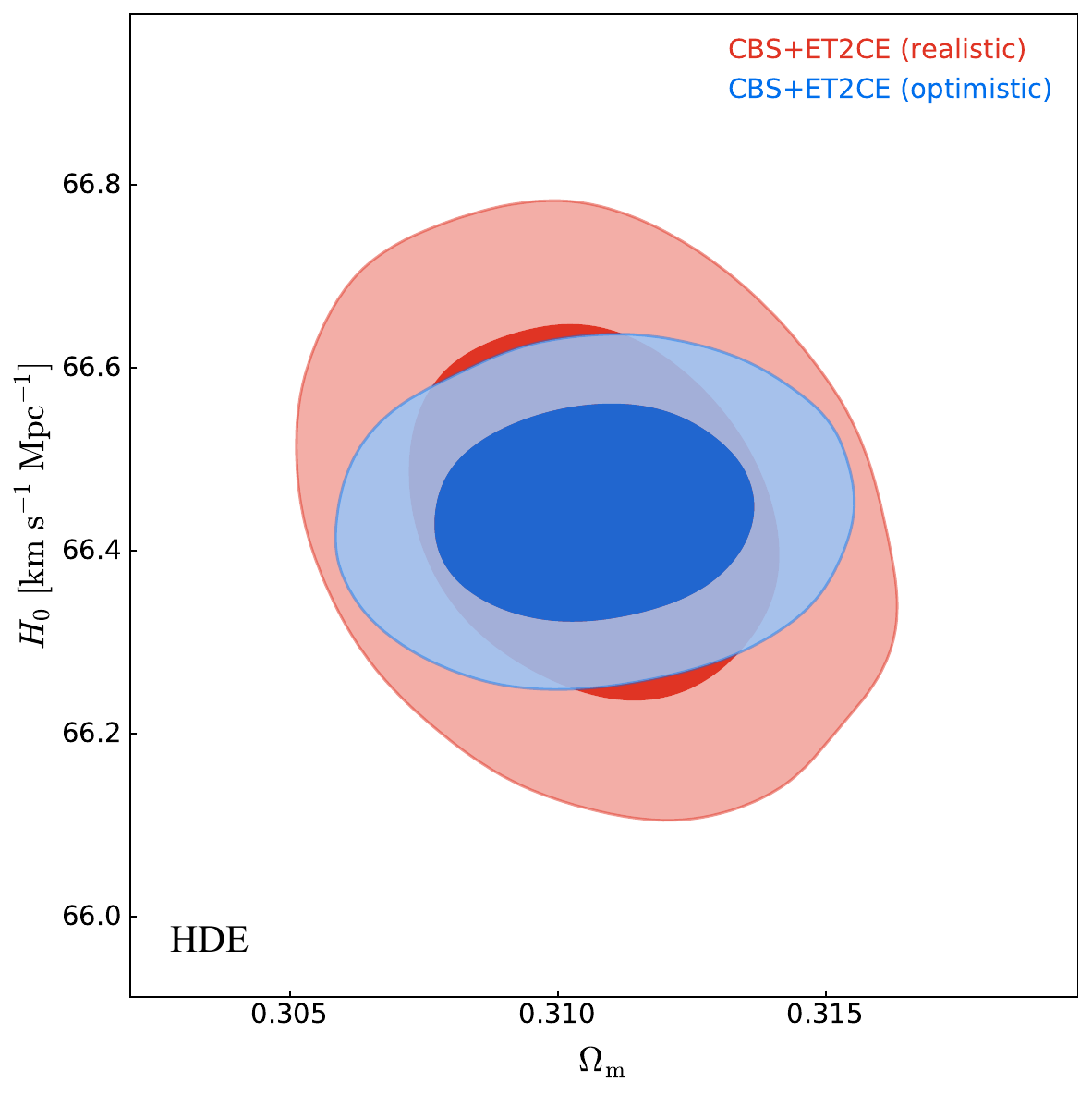}
	\includegraphics[width=0.45\linewidth,angle=0]{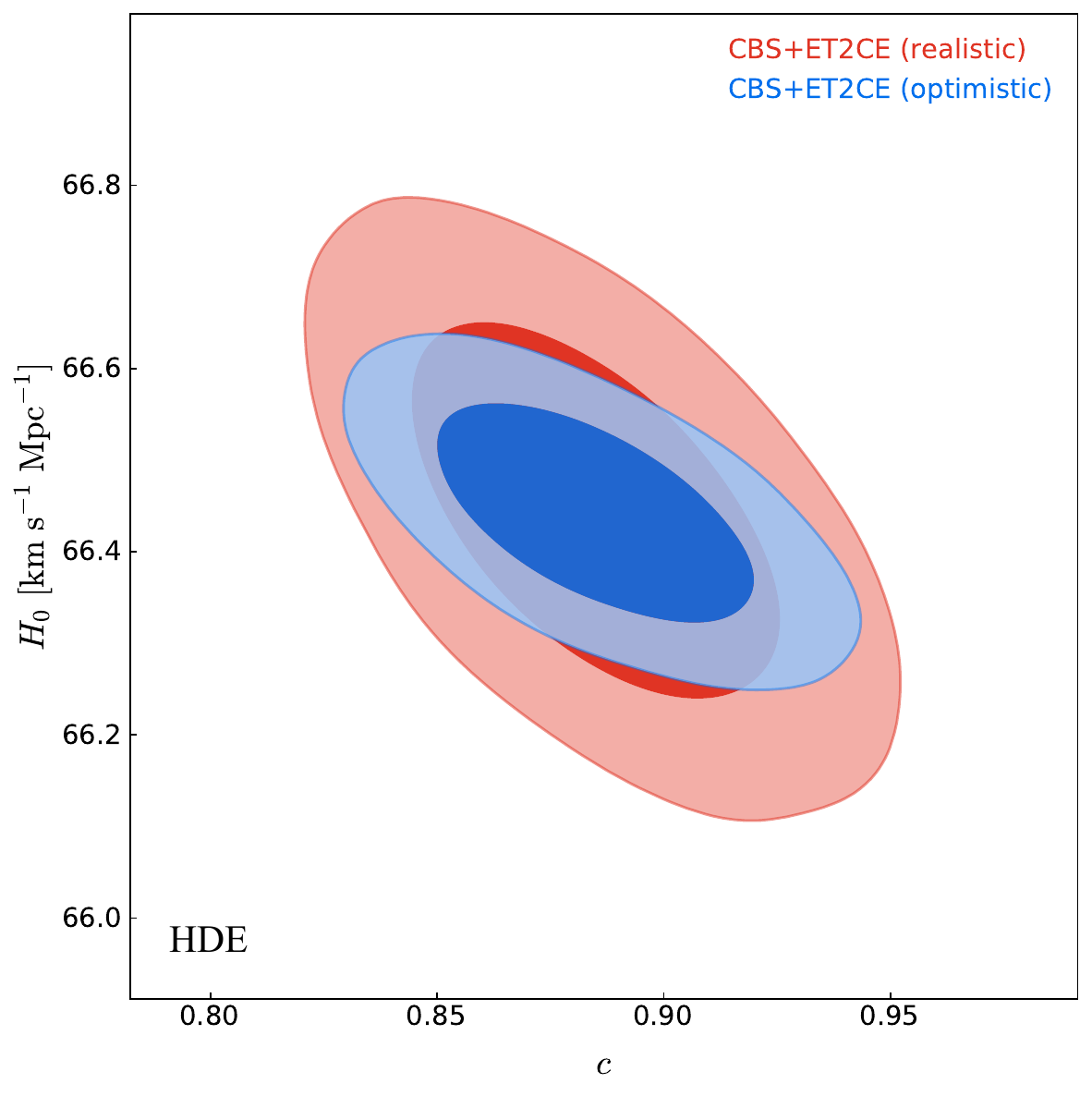}
	\caption{\label{fig9} Two-dimensional 
 marginalized contours ($68.3\%$ and $95.4\%$ confidence level) in the $\Omega_{\rm m}$--$H_0$ and $c$--$H_0$ planes for the HDE model in the realistic and optimistic scenarios of the CBS + ET2CE data.}
\end{figure}

\begin{figure}[H]
	\includegraphics[width=0.45\linewidth,angle=0]{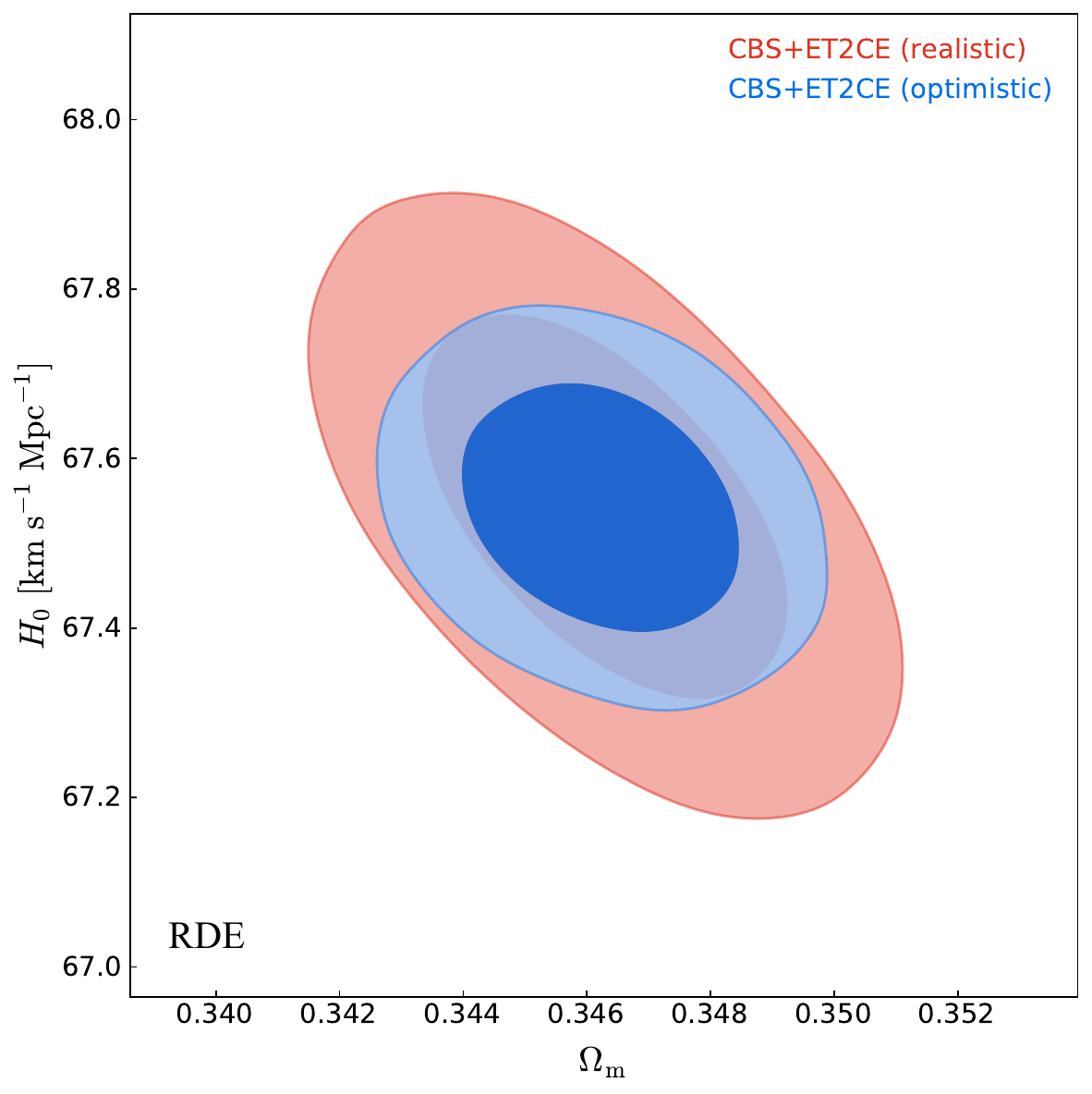}
	\includegraphics[width=0.45\linewidth,angle=0]{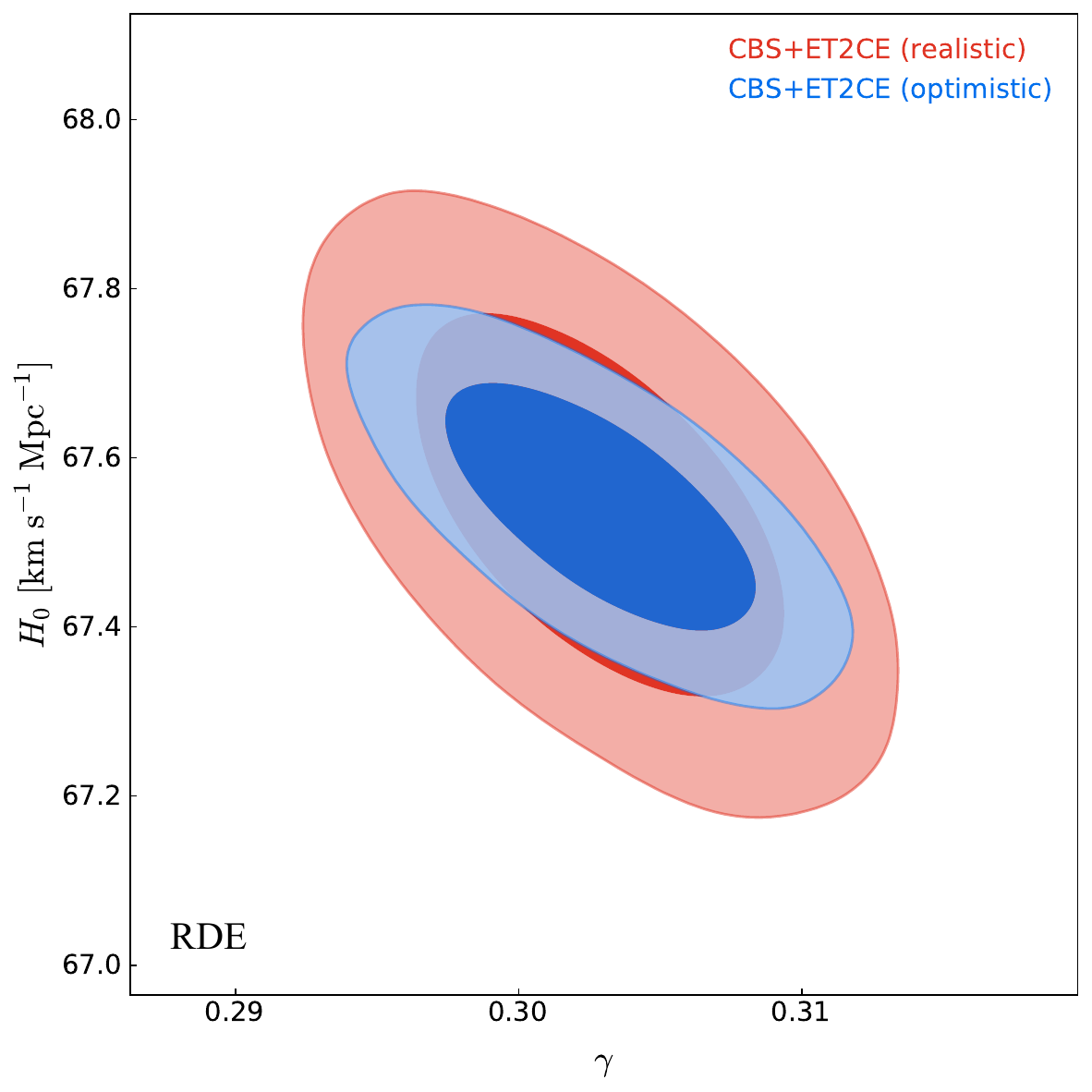}
	\caption{\label{fig10} Two-dimensional 
 marginalized contours ($68.3\%$ and $95.4\%$ confidence level) in the $\Omega_{\rm m}$--$H_0$ and $\gamma$--$H_0$ planes for the RDE model in the realistic and optimistic scenarios of the CBS + ET2CE data.}
\end{figure}

Subsequently, we present the main constraint results in Figures~\ref{fig5}--\ref{fig10} and Tables~\ref{tab2} and~\ref{tab3}. The 1$\sigma$ and 2$\sigma$ posterior distribution contours for the relevant cosmological parameters are shown in Figures~\ref{fig5}--\ref{fig10} and the 1$\sigma$ errors for the marginalized parameter constraints are shown in Tables~\ref{tab2} and~\ref{tab3}. We adopt $\sigma(\xi)$ and $\varepsilon(\xi)$ to represent the absolute and the relative errors of parameter $\xi$, with $\varepsilon(\xi)=\sigma(\xi)/\xi$. In the following, we use ET2CE as the representative GW data for some relevant discussions, as it achieves the highest measurement precision for $d_{\rm L}$.

\vspace{-12pt} 

\vspace{-12pt} 
\vspace{-12pt} 
\vspace{-12pt} 

\vspace{-6pt} 
\begin{table}[H]
	\caption{The 
 absolute (1$\sigma$) and relative errors of the cosmological parameters in the HDE and RDE models using the ET, CE, 2CE, ET2CE, CBS, CBS + ET, CBS + CE, CBS + 2CE, and CBS + ET2CE data in the optimistic scenario. $H_0$ is in units of $\rm km\ s^{-1}\ Mpc^{-1}$.}
	\label{tab2}
	
	\begin{adjustwidth}{-\extralength}{0cm}
		\setlength{\tabcolsep}{1.8mm}
		\renewcommand{\arraystretch}{1}
		{
			\begin{tabular}{{ccccccccccc}}
				\toprule
%
				\textbf{Model}      & \textbf{Error}             &\textbf{ET}          &\textbf{CE}          &\textbf{2CE}          &\textbf{ET2CE }        &\textbf{CBS}         &\textbf{CBS + ET}      &\textbf{CBS + CE}     &\textbf{CBS + 2CE}      &\textbf{CBS + ET2CE} \\ \midrule
				\multirow{6}{*}{HDE}
				&$\sigma(\Omega_{\rm m})$          &$0.0680$  &$0.0430$  &$0.0305$&$0.0250$  &$0.0067$  &$0.0024$  &$0.0022$  &$0.0020$  &$0.0020$\\
				&$\sigma(H_0)$                               &$0.295$  &$0.170$  &$0.130$  &$0.120$  &$0.680$  &$0.170$  &$0.100$  &$0.086$  &$0.079$         \\
				&$\sigma(c)$                           &$0.370$  &$0.200$  &$0.140$  &$0.114$  &$0.041$  &$0.028$  &$0.025$  &$0.024$  &$0.023$      \\
				&$\ve(\Omega_{\rm m})$             &$25.00\%$  &$14.58\%$  &$10.07\%$  &$8.17\%$  &$2.16\%$  &$0.77\%$  &$0.71\%$  &$0.64\%$  &$0.64\%$   \\
				&$\ve(H_0)$                                   &$0.44\%$  &$0.26\%$  &$0.20\%$  &$0.18\%$  &$1.02\%$  &$0.26\%$  &$0.15\%$  &$0.13\%$   &$0.12\%$   \\
				&$\ve(c)$                                   &$32.74\%$  &$20.41\%$  &$15.05\%$  &$12.46\%$  &$4.62\%$  &$3.16\%$  &$2.82\%$  &$2.71\%$   &$2.60\%$     \\ \midrule
				\multirow{6}{*}{RDE}
				&$\sigma(\Omega_{\rm m})$       &$0.0285$  &$0.0180$  &$0.0140$  &$0.0110$  &$0.0074$  &$0.0023$  &$0.0018$  &$0.0015$   &$0.0015$   \\
				&$\sigma(H_0)$                           &$0.390$  &$0.200$  &$0.160$  &$0.140$  &$0.590$  &$0.190$  &$0.120$  &$0.100$   &$0.097$    \\
				&$\sigma(\gamma)$                     &$0.0300$  &$0.0190$  &$0.0140$  &$0.0120$  &$0.0050$  &$0.0045$  &$0.0042$  &$0.0039$   &$0.0036$  \\
				&$\ve(\Omega_{\rm m})$            &$8.33\%$  &$5.22\%$  &$4.05\%$  &$3.18\%$  &$2.14\%$  &$0.66\%$  &$0.52\%$  &$0.43\%$   &$0.43\%$  \\
				&$\ve(H_0)$                                  &$0.58\%$  &$0.30\%$  &$0.24\%$  &$0.21\%$  &$0.87\%$  &$0.28\%$  &$0.18\%$  &$0.15\%$   &$0.14\%$   \\
				&$\ve(\gamma)$                            &$9.74\%$  &$6.25\%$  &$4.62\%$  &$3.96\%$  &$1.65\%$  &$1.49\%$  &$1.39\%$  &$1.29\%$   &$1.19\%$     \\  \bottomrule				
		\end{tabular}}
	\end{adjustwidth}
\end{table}
\vspace{-12pt} 
\begin{table}[H]
	\caption{Same 
 as in Table~\ref{tab2}, but assuming the realistic scenario.}
	\label{tab3}
	\setlength{\tabcolsep}{1.72mm}
	\renewcommand{\arraystretch}{1}
	\begin{adjustwidth}{-\extralength}{0cm}
		{
			\begin{tabular}{{ccccccccccc}}
				\toprule
%
				\textbf{Model}      & \textbf{Error}             &\textbf{ET}          &\textbf{CE}          &\textbf{2CE  }        &\textbf{ET2CE    }     &\textbf{CBS }        &\textbf{CBS + ET }     &\textbf{CBS + CE }    &\textbf{CBS + 2CE }     &\textbf{CBS + ET2CE} \\ \midrule
				\multirow{6}{*}{HDE}
				&$\sigma(\Omega_{\rm m})$          &$0.0920$&$0.0690$  &$0.0550$  &$0.0445$  &$0.0067$  &$0.0030$  &$0.0028$  &$0.0025$  &$0.0023$      \\
				&$\sigma(H_0)$                               &$0.43$  &$0.34$  &$0.27$  &$0.24$  &$0.68$  &$0.25$  &$0.21$  &$0.16$  &$0.14$         \\
				&$\sigma(c)$                           &$0.490$  &$0.380$  &$0.280$  &$0.220$  &$0.041$  &$0.030$  &$0.028$  &$0.028$  &$0.027$     \\
				&$\ve(\Omega_{\rm m})$             &$34.33\%$&$24.73\%$  &$19.10\%$  &$15.08\%$  &$2.16\%$  &$0.97\%$  &$0.90\%$  &$0.80\%$  &$0.74\%$   \\
				&$\ve(H_0)$                                   &$0.64\%$  &$0.51\%$  &$0.41\%$  &$0.36\%$  &$1.02\%$  &$0.38\%$  &$0.32\%$  &$0.24\%$  &$0.21\%$   \\
				&$\ve(c)$                                   &$41.53\%$  &$33.93\%$  &$26.67\%$  &$22.22\%$  &$4.62\%$  &$3.39\%$  &$3.17\%$  &$3.16\%$  &$3.05\%$     \\ \midrule
				\multirow{6}{*}{RDE}
				&$\sigma(\Omega_{\rm m})$       &$0.0535$  &$0.0345$  &$0.0245$  &$0.0200$  &$0.0074$  &$0.0032$  &$0.0027$  &$0.0021$  &$0.0019$      \\
				&$\sigma(H_0)$                           &$0.61$  &$0.42$&$0.31$  &$0.27$  &$0.59$  &$0.27$  &$0.21$  &$0.16$  &$0.15$       \\
				&$\sigma(\gamma)$                      &$0.0520$  &$0.0375$  &$0.0260$  &$0.0220$  &$0.0050$  &$0.0048$  &$0.0047$  &$0.0045$  &$0.0043$        \\
				&$\ve(\Omega_{\rm m})$            &$15.92\%$  &$10.09\%$  &$7.12\%$  &$5.78\%$  &$2.14\%$  &$0.92\%$  &$0.78\%$  &$0.61\%$  &$0.55\%$  \\
				&$\ve(H_0)$                                  &$0.90\%$  &$0.61\%$  &$0.46\%$  &$0.40\%$  &$0.87\%$  &$0.40\%$  &$0.31\%$  &$0.24\%$  &$0.22\%$   \\
				&$\ve(\gamma)$                           &$16.51\%$  &$12.18\%$  &$8.52\%$  &$7.24\%$  &$1.65\%$  &$1.58\%$  &$1.55\%$  &$1.49\%$  &$1.42\%$     \\  \bottomrule				
		\end{tabular}}
	\end{adjustwidth}
\end{table}

In this paper, the primary objective is to assess the impact of joint observations between 3G GW detectors and future GRB detectors on cosmological measurements in the HDE and RDE models, as well as their ability to break the degeneracies in cosmological parameters commonly observed in traditional EM data. To illustrate this, we have performed simulations to generate mock GW data, which we have subsequently combined with mainstream EM observations, i.e., CBS data. Our analysis focuses on the estimation errors and the precision of cosmological parameters derived from this combined dataset. To simplify the calculations, we only adopt the ``Planck distance priors’’ from the Planck 2018 observation as the CMB observation in our calculations since the ``distance priors'' and the CMB power spectrum provide similar constraints~\cite{Li:2012spm}. Although combining the CMB power spectrum along with the DESI 2024 BAO~\cite{DESI:2024uvr,DESI:2024lzq} and eBOSS DR16~\cite{eBOSS:2019dcv} data would slightly enhance the precision of cosmological parameters, it would not significantly affect the ability of GW data to break the degeneracies in cosmological parameters observed in CBS data.

\subsection{Constraint Results in Optimistic Scenario}

In Figure~\ref{fig5}, we show the constraint results for the HDE model. It is clear that ET2CE provides the tightest constraints, followed by 2CE, CE, and ET. The primary reason is that the constraint results are highly influenced by the numbers and errors of the standard siren data, while ET2CE offers the highest number of data points with the lowest errors, followed by 2CE, CE, and ET. Nonetheless, ET gives $\sigma(H_0)=0.295~\rm km~s^{-1}~Mpc^{-1}$ with a constraint precision of $0.44\%$, which is much better than that of CBS. However, for other cosmological parameters, the GW data alone can only provide rather weak measurements. Fortunately, GW can effectively break the cosmological parameter degeneracies generated by CBS, as shown in Figure~\ref{fig6}. As can be seen, the parameter degeneracy orientations of ET2CE and CBS are different. With the addition of ET2CE to CBS, the constraint precision of cosmological parameters is significantly improved. CBS + ET2CE gives $\sigma(c)=0.0023$ and $\sigma(H_0)=0.079~\rm km~s^{-1}~Mpc^{-1}$, which are $43.9\%$ and $88.4\%$ better than those of CBS. In addition, the constraint precisions of $\Omega_{\rm m}$ and $H_0$ are $0.64\%$ and $0.12\%$, both of which are significantly better than $1\%$, the standard of precision cosmology. Meanwhile, from Figure~\ref{fig6}, we can also find that the central value of $c$ given by CBS is 0.887. Since $c<1$, the EoS of dark energy in the HDE model crosses the phantom divide of $w=-1$, which means that the current and future universe is dominated by phantom energy and will end up with a ``big rip'' singularity. Since the GW data are simulated based on the CBS constraints, the combination of CBS and GW cannot affect the central value of $c$, but it can reduce the error of $c$. For the same reason, this paper does not address the consistency between GW and CBS (including the latest DESI BAO) constraints.

In Figure~\ref{fig7}, we show the constraint results for the RDE model. The constraint results for GW still hold, with ET2CE providing the tightest constraints, followed by 2CE, CE, and ET. For the constraint precision of the parameters $\Omega_{\rm m}$, ET2CE gives $\Omega_{\rm m}=0.0110$, which is slightly worse than that of CBS. With the addition of ET2CE to CBS, the constraints on $\gamma$ and $H_0$ could be improved by $28.0\%$ and $83.6\%$. Meanwhile, CBS + ET2CE gives $\sigma(\gamma)=0.0036$ with a precision of $1.19\%$, which is nearly at the level expected in precision cosmology. It is promising that the fundamental nature of dark energy could be explored using 3G GW standard sirens.

\subsection{Constraint Results in Realistic Scenario}

In Figures~\ref{fig8} and~\ref{fig9}, we present the constraint results for the HDE model. In Figure~\ref{fig8}, we can clearly find that ET2CE (optimistic) gives tighter constraint results than those of ET2CE (realistic). ET2CE (realistic) givens $\sigma(c)=0.220$ and $\sigma(H_0)=0.24$, which are $93.0\%$ and $100.0\%$ worse than those of ET2CE (optimistic). In Figure~\ref{fig9}, we can find that CBS~+~ET2CE (optimistic) also gives tighter constraints on cosmological parameters than those of CBS~+~ET2CE (realistic). CBS + ET2CE (realistic) givens $\sigma(c)=0.027$ and $\sigma(H_0)=0.14$, which are $17.4\%$ and $77.2\%$ worse than those of CBS + ET2CE (optimistic). In Figure~\ref{fig10}, we show the case of the RDE model. We can also find that CBS + ET2CE (optimistic) gives tighter constraints on cosmological parameters than those of CBS + ET2CE (realistic). CBS + ET2CE (realistic) givens $\sigma(\gamma)=0.0043$ and $\sigma(H_0)=0.15$, which are $19.4\%$ and $54.6\%$ worse than those of CBS + ET2CE (optimistic).

\subsection{{Methodological Improvements Compared to Previous Work}}

Compared to previous work~\cite{Zhang:2019ple}, the main differences in our methodological improvements are as follows.

In previous paper, it was commonly assumed that there will be around 1000 standard sirens with detectable EM counterparts for either ET or CE alone over a 10-year observation. However, this assumption is overly optimistic and may not be realistic. For instance, despite the optimistic estimate for a single ET detector, we expect only about 252 detectable GW-GRB events. Even with the ET2CE network, the number of detectable GW-GRB events is 363. This clearly indicates that the number of standard sirens has been considerably overestimated in previous paper.

Previous work typically assumed that the redshift distribution of standard sirens directly followed the one derived from the star formation rate, which had long tails at higher redshifts. However, in joint GW-GRB observations, the detection rates of both GW and GRB detectors for events across different redshifts must be taken into account. As a result, the redshift distribution in our analysis is shifted to lower values, primarily within the range $z\in [0,2]$, as shown in Figures~\ref{fig1} and~\ref{fig2}, rather than extending up to $z\in [0,5]$.

In previous paper, the influence of Earth's rotation was overlooked. However, considering the significant impact of this effect on 3G GW detectors~\cite{Han:2023exn}, we incorporate Earth's rotational effects in our simulation of GW standard sirens to more accurately represent real observational conditions.

In this work, we use the updated ``Pantheon+'' compilation~\cite{Brout:2022vxf}, which includes 1701 light curves from 1550 distinct objects, compared to the earlier ``Pantheon'' dataset with 1048 data points~\cite{Pan-STARRS1:2017jku}. This upgrade significantly improves the constraining power of Pantheon+ relative to the original Pantheon dataset.

\subsection{{Constraint Results Compared to Previous Works}}

In Ref.~\cite{Zhang:2019ple} for ET alone in the HDE model, they gave $\sigma(H_0)=0.59~\rm km~s^{-1}~Mpc^{-1}$, which is worse than our constraint result under the realistic scenario. However, for the parameter $\Omega_{\rm m}$, they gave $\sigma(\Omega_{\rm m})=0.0306$, which is tighter than our result under the optimistic scenario. This is primarily due to the fact that, compared to previous work, the redshift distributions of GW standard sirens in our analysis are lower, as mentioned above. In the early universe, the expansion history is dominated by $\Omega_{\rm m}$, which makes low-redshift GW events less effective at constraining $\Omega_{\rm m}$. On the other hand, low-redshift GW events primarily capture the direct effects of the Hubble constant $H_0$, thus providing stronger constraints on $H_0$. For the parameter $c$, they gave $\sigma(c)=0.218$, which is tighter than our optimistic scenario result. Similar conclusions generally hold for the RDE model. It should be noted that, when combined with CBS, our constraint results under the realistic scenario provide better or slightly better constraints than those presented in their analysis. This is primarily because, for SN data, we utilized the updated ``Pantheon+'' dataset, which provides stronger constraints than the ``Pantheon'' dataset used in their analysis.

Then, we turn our attention to the $w$CDM and I$\Lambda$CDM models, which also contain just one extra parameter compared to the $\Lambda$CDM model. In our previous work~\cite{Han:2023exn}, we employed a similar methodology to obtain the constraint results for both the $w$CDM and I$\Lambda$CDM ($Q=\beta H \rho_c$) models. For ET alone in the $w$CDM model, we obtained $\sigma(w) = 0.120$ under the optimistic scenario, which is slightly tighter than the constraint results provided by Zhang et al.
~\cite{Zhang:2019ple}. In the I$\Lambda$CDM model, under the realistic scenario with CBS + ET, we obtained $\sigma(\beta) = 0.00081$, which is better than the result given by Li {et al}.~\cite{Li:2019ajo}, who adopted a methodology similar to Zhang {et al}.~\cite{Zhang:2019ple}.

\subsection{{{Challenges of the HDE Model}}}

{Recently, Li {et al}.~\cite{Li:2024qus} investigated the cosmological implications of the HDE model using CMB power spectrum, DESI 2024 BAO, and SN data. They found that when evaluated with DESI BAO and SN data, the HDE model performs comparably to the $\Lambda$CDM model based on the Bayesian evidence. However, the inclusion of CMB data makes the HDE model significantly less favored than the $\Lambda$CDM model, although this is not sufficient to definitively refute the HDE model. In addition, the systematic errors in CMB observation (e.g., the measurement of temperature or polarization) may also affect this conclusion. Therefore, despite challenges posed by current observations, the HDE model has not been definitively excluded. It merits further investigation, particularly with the availability of more precise late-universe observations in the future, especially GW observations, which could offer critical insights into the nature of dark energy.}

{In addition, the holographic principle may also encounter observational challenges. It is noted that quantum fluctuations in spacetime would lead to apparent ``holographic noise'', which is measurable by GW detectors~\cite{Hogan:2007pk}. The GEO600 detector observed unexplained ``mysterious noise'' in its most sensitive frequency range of 300--1400 Hz, with a spectrum roughly consistent with the predictions of holographic noise~\cite{Hogan:2008zw}. However, analyses in 2011 of measurements of GRB 041219A failed to detect holographic noise at the expected scale~\cite{ESA}. Despite these challenges, the holographic principle remains an intriguing area of investigation, as it offers a potential framework for understanding quantum gravity and could provide new insights in the future with more refined experiments and observations.}

\section{Conclusions}\label{sec5}

In this paper, we investigate the ability of 3G GW detectors to constrain HDE and RDE models using GW standard sirens. We examine the synergy between 3G GW detectors and the THESEUS-like GRB detector for multi-messenger observations. Specifically, we evaluate four detection strategies: ET, CE, the 2CE network, and the ET2CE network. Additionally, we consider both optimistic (all detected short GRBs can determine redshifts) and realistic (only 1/3 of detected short GRBs can determine redshifts) scenarios for the FoV in order to conduct the multi-messenger analysis.

We find that GW data alone can provide tight constraints on $H_0$, with precision of 0.18--0.64\% in the HDE model. However, it offers loose constraints on other cosmological parameters. Fortunately, due to its ability to break parameter degeneracies generated by other EM observations, GW can play a vital role in improving the overall parameter estimation. With the addition of GW to CBS, the constraints on cosmological parameters $H_0$, $c$ and $\Omega_{\rm{m}}$ can be improved by $63.2$--$88.4\%$, $26.8$--$43.9\%$ and $55.2$--$70.1\%$ in the HDE model. Additionally, although current observations have already excluded the RDE model, the use of GW data will further enhance the accuracy of parameter estimation for this model. 

\vspace{+6pt} 

\authorcontributions{Conceptualization, J.-F.Z. and X.Z.; methodology, T.H.; software, T.H. and Z.L.; validation, T.H. and Z.L.; formal analysis, J.-F.Z.; investigation, Z.L.; writing—original draft preparation, Z.L. and T.H.; writing—review and editing, T.H. and Z.L.; supervision, J.-F.Z. and X.Z.; project administration, X.Z. All authors have read and agreed to the published version of the manuscript.}

\funding{This research was funded by the National SKA Program of China (Grants Nos. 2022SKA0110200 and 2022SKA0110203), the National Natural Science Foundation of China (Grants Nos. 12473001, 11975072, and 11875102), and the 111 Project (Grant No. B16009).}

\dataavailability{Not applicable.}

\acknowledgments{We 
 acknowledge the support of the National SKA Program of China (Grants Nos. 2022SKA0110200 and 2022SKA0110203), the National Natural Science Foundation of China (Grants Nos. 12473001, 11975072, and 11875102), and the 111 Project (Grant No. B16009).}

\conflictsofinterest{The authors declare no conflict of interest.}

\begin{adjustwidth}{-\extralength}{0cm}

\reftitle{References}

\PublishersNote{}
\end{adjustwidth}
\end{document}